\documentclass[12pt,reqno]{amsart}
\usepackage{amsaddr}
\usepackage{graphicx,amssymb,amsmath,amsthm}

\usepackage{natbib}
\usepackage[colorlinks,bookmarksopen,bookmarksnumbered,citecolor=red,urlcolor=red]{hyperref}

\newcommand\BibTeX{{\rmfamily B\kern-.05em \textsc{i\kern-.025em b}\kern-.08em
		T\kern-.1667em\lower.7ex\hbox{E}\kern-.125emX}}

\usepackage{amssymb}
\usepackage{bm,graphicx,textcomp,bbm} % Math packagef\boldsymbol

\usepackage{subcaption}
\usepackage{moreverb}
\usepackage[colorinlistoftodos]{todonotes}

\newcounter{bgcomment}

\newcounter{gagcomment}

\def\deltamin{\delta_{\rm{min}}}

\title{Stochastically perturbed bred vectors}

\author{Brent Giggins and Georg A. Gottwald}

\address{School of Mathematics and Statistics, University of Sydney, Australia}
\email[B. Giggins and G. A. Gottwald]{brent.giggins@sydney.edu.au {\rmfamily and} georg.gottwald@sydney.edu.au}

\begin{document}
	
%%%%%%%%%%%%%%%%%%%%%%%%%%%%%%%%%%%%%%%%%%%%%%%%%%%%%%%%%%%%%%%%%%%%%

%%%%%%%%%%%%%%%%%%%%%%%%%%%%%%%%%%%%%%%%%%%%%%%%%%%%%%%%%%%%%%%%%%%%%
% ABSTRACT

\begin{abstract}
The breeding method is a computationally cheap procedure to generate initial conditions for ensemble forecasting which project onto relevant synoptic growing modes. Ensembles of bred vectors, however, often lack diversity and align with the leading Lyapunov vector, which severely impacts their statistical reliability. In previous work we developed stochastically perturbed bred vectors (SPBVs) and random draw bred vectors (RDBVs) in the context of multi-scale systems. Here we explore when this method can be extended to systems without scale separation, and examine the performance of the stochastically modified bred vectors in the single scale Lorenz 96 model. In particular, we show that the performance of SPBVs crucially depends on the degree of localisation of the bred vectors.  It is found that, contrary to the case of multi-scale systems, localisation is detrimental for applications of SPBVs in systems without scale-separation when initialised from assimilated data. In the case of weakly localised bred vectors, however, ensembles of SPBVs constitute a reliable ensemble with improved ensemble forecasting skills compared to classical bred vectors, while still preserving the low computational cost of the breeding method. RDBVs are shown to have superior forecast skill and form a reliable ensemble in weakly localised situations, but in situations when they are strongly localised they do not constitute a reliable ensemble and are over-dispersive.
\end{abstract}

%\keywords{ensemble forecasting; ensemble methods; bred vectors; covariant Lyapunov vectors;}
	
\maketitle
		
%%%%%%%%%%%%%%%%%%%%%%%%%%%%%%%%%%%%%%%%%%%%%%%%%%%%%%%%%%%%%%%%%%%%%
% MAIN BODY OF PAPER
%%%%%%%%%%%%%%%%%%%%%%%%%%%%%%%%%%%%%%%%%%%%%%%%%%%%%%%%%%%%%%%%%%%%%

\section{{\bf{Introduction}}}

The chaotic nature of the atmosphere and the climate system, and its sensitivity to small uncertainties in the initial conditions may render single forecasts meaningless. Probabilistic forecasts, which instead are derived from an ensemble of forecasts, have become standard in numerical weather forecasting, providing a Monte-Carlo estimate of the probability density function \citep{Epstein69,Leith74,LeutbecherPalmer08}. Such ensemble forecasts issue the most probable forecast alongside measures of its uncertainty. A key question is how to initialise the ensemble members. There exist several methods to generate such ensembles, using singular vectors \citep{Lorenz65,Palmer93}, bred vectors \citep{TothKalnay93,TothKalnay97}, analysis ensembles from ensemble Kalman filters \citep{Evensen94,HoutekamerMitchell98,WangBishop03,BuizzaEtAl05}, and more recently model generated analogs \citep{AtenciaZawadzki17}. In this work we consider bred vectors and the so called "breeding method" which constitutes a computationally very attractive method to produce an ensemble of initial conditions introduced by \cite{TothKalnay93,TothKalnay97}. In this method initial conditions are generated from finite perturbations, the bred vectors (BVs), which encapsulate information about fast growing modes. Such fast growing initial conditions are then likely to be pre-images of states of high probability. Bred vectors have been successfully implemented for more than a decade since 1992 by the National Centre for Environmental Prediction (NCEP) for their operational 1-15 day ensemble forecasts. Applications range from ENSO prediction \citep{CaiEtAl03,ChengEtAl10}, seasonal-to-interannual forecasting in coupled general circulation models (CGCMs) \citep{YangEtAl09} to forecasting weather and climate on Mars \citep{NewmanEtAl04,GreybushEtAl13}.\\

In the breeding method a control trajectory alongside an ensemble of nearby trajectories is generated. The ensemble members are initialised from perturbed initial conditions with finite perturbation size $\delta$ from the initial condition of the control trajectory. Different as for Lyapunov vectors, all ensemble members are propagated with the full nonlinear model. The perturbed trajectories are periodically rescaled to a specified finite-size distance $\delta$ away from the control trajectory, to avoid saturation of instabilities. Bred vectors are defined as the difference at the time of rescaling of the perturbed trajectories and the control trajectory. The perturbation size is often thought of as a a filter of small scale instabilities, in the sense that BVs are insensitive to very fast growing instabilities which typically are associated with small scale processes and which nonlinearly saturate at an amplitude smaller than $\delta$. Choosing $\delta$ appropriately allows the forecast to be tuned to specific instabilities of interest. For perturbation sizes of the order of $1-10\%$ of the natural variability in the atmosphere, BVs were found to significantly project onto baroclinic instabilities  \citep{TothKalnay97,CorazzaEtAl03}. Moreover, bred vectors allow for the prediction of regime changes for perturbation sizes in a certain range \citep{PenaEtAl04,EvansEtAl04,NorwoodEtAl13}).\\

Ideally bred vectors constitute a sufficiently diverse ensemble mediated by the provided stochasticity of saturated sub-synoptic processes \citep{TothKalnay97}. Bred vector ensembles, however, may lack diversity in certain situations and most of the ensemble forecast variability may be captured by a single BV \citep{WangBishop03}. Indeed, for a range of small perturbation sizes $\delta$ bred vectors align with the leading Lyapunov vector and the ensemble collapses to a single member. When this occurs, this reduction in ensemble diversity is hugely detrimental for the ability of a bred vector ensemble to reliably sample the forecast probability density function. To preserve the attractive features of BVs such as their low computational cost, several methods were proposed to increase the ensemble spread in BVs. In particular by orthogonalizing bred vectors \citep{Annan04,KellerEtAl10}, by introducing stochasticity either via small random perturbations at each rescaling period \citep{GreybushEtAl13} or via stochastic backscattering \citep{OKaneFrederiksen08}, by rescaling using the geometric rather than the Euclidean norm \citep{PrimoEtAl08,PazoEtAl11,PazoEtAl13} or by changing the rescaling procedure based on the largest BV 
\citep{BalciEtAl12}. \\

In recent work, \cite{GigginsGottwald19} proposed a method of stochastically perturbing BVs in the context of ensemble forecasts of the slow dynamics in multi-scale systems to alleviate the problem of small ensemble diversity, introducing stochastically perturbed bred vectors (SPBV). SPBVs were constructed, it was argued, to sample the conditional probability function of the system conditioned on the slow variables by multiplicatively randomising the fast BV components. The localised character of the fast BV components ensures that the perturbed initial conditions relax after a short transient to initial conditions which are close to those of their parent BV, with slow components being close in phase space to those of the original BV and with fast components being dynamically conditioned on the slow components. Hence, these stochastically perturbed BVs generate initial conditions which sample the probability density function conditioned on the slow synoptic state. SPBVs exhibit a markedly increased ensemble dimension, in particular for small but finite perturbation sizes. It was shown that the subsynoptic variability associated with SPBVs generated synoptic variability of the same order as suggested by the analysis fields. The increased diversity of SPBVs lead to an improved forecasting skill when compared to standard BVs. Important for probabilistic forecasts, SPBV ensembles were shown to be reliable in the sense that each ensemble member is equally likely to be closest to the truth. Furthermore, SPBVs were shown to be dynamically consistent and recover characteristic features of the temporal evolution of errors in chaotic dynamical systems. Additionally, random draw bred vectors (RDBVs), which are designed to sample from the marginal equilibrium density of the fast variables (and hence are not conditioned on the slow variables), were introduced. While RDBVs are not dynamically consistent and are typically over-dispersive, they were found to still have improved forecast skill over standard BVs.\\

In this work we explore if, and under which conditions, the ideas proposed in \cite{GigginsGottwald19} for multi-scale dynamics can be applied to the situation of general dynamical systems without time-scale separation. Single-scaled models such as the quasi-geostrophic equations are often used to study the slow large-scale dynamics of the atmosphere-ocean system. Filtering out fast small-scale processes has the computational advantage of avoiding the numerical difficulties associated with multi-scale systems. We study here in what way stochastically modifying classical BVs may help in using the attractive features of BVs, such as their low computational coast and their dynamic adaptivity in the sense that they resemble realistic error growth. In the realistic situation when the state of the atmosphere is given by the analysis output from a data assimilation procedure, a good forecast ensemble has to satisfy two constraints: It has to evolve into likely future states, and it has to account for the uncertainty of the analysis used to generate the ensemble. In the case of multi-scale dynamics these issues were resolved by generating the necessary small synoptic uncertainty required by the analysis covariance via stochastically perturbing the fast variables, which subsequently quickly relaxed onto the attractor. The situation in single-scale dynamics is more complicated. Whereas localisation of the bred vectors was beneficial in the multi-scale case and allowed for the conditioning of the SPBVs on the slow synoptic dynamical state, localisation of BVs prohibits in the single-scale scenario perturbations outside the localised region. Hence, although the resulting perturbation will be close in phase space to the original BV and appropriately sample the probability density function around it, the resulting initial conditions may not contain sufficient variability in the regions of significant uncertainty of the analysis.\\  %Here, the key to generating SPBVs relies on preserving the spatial correlations of the perturbation patterns, corresponding to a local in space error growth. Stochastically perturbing these perturbations then creates initial conditions which are off the attractor. Invoking fast relaxation towards the attractor, these stochastically SPBVs will lie close in phase space to their parent BV and exhibit a similar localization pattern. As such SPBVs and their subsequent relaxation towards the attractor effectively sample the probability density function conditioned on the current state of the system. We will show that the strength of the multiplicative noise used for generating SPBV ensembles plays an important role, where a trade-off in how strong the multiplicative noise $\sigma$ of an SPBV perturbation has to be made in order to balance between providing the most ensemble diversity while preserving some of the spatial correlation. 

To investigate the performance of stochastically modified bred vectors in a single-scale system we consider the Lorenz 96 model \citep{Lorenz96} in two settings, which support strongly localised and weakly localised BVs. Our numerical simulations demonstrate that in both cases SPBVs and RDBVs exhibit significantly increased diversity and ensemble dimension, and that in both cases they provide superior forecast skill and reliability measures when compared to classical BVs. Their forecast skill and reliability, diagnosed by means of error-spread relationships and reliability diagrams, however, crucially depends on the degree of localisation, when the ensemble is centred around an analysed field. In the weakly localised case SPBVs and RDBVs constitute reliable forecast ensembles with forecast skills comparable to a reference ensemble obtained from an ensemble transform Kalman filter (ETKF). In the case of strongly localised bred vectors, however, their reliability is severely impeded. Localisation prevents SPBVs to constitute a reliable ensemble as they are not consistent with the analysis error which may be non-negligible outside the region of significant activity of the SPBV. For the weakly localised case we will show that the strength of the multiplicative noise used for generating SPBV ensembles can be judiciously chosen as a trade-off between providing the most ensemble diversity while preserving dynamic adaptivity in the sense that they resemble realistic error growth comparable to those of Lyapunov vectors. This dynamical consistency is probed by projecting onto the subspace spanned by the dominant covariant Lyapunov vectors. RDBVs will be shown to be dynamically inconsistent but to nevertheless feature improved forecast skill over SPBVs.\\

The paper is organised as follows. In Section~\ref{sec.model} we introduce the Lorenz 96 model \citep{Lorenz96}. In Section~\ref{sec.BV} we briefly review the breeding method. Section~\ref{sec.SPBV} introduces our stochastically modified bred vector ensemble methods, namely SPBVs and RDBVs, and shows how they relate to covariant Lyapunov modes as a measure of their dynamic adaptivity. Section~\ref{sec.diagnostics} introduces the diagnostics used to evaluate the performance and the efficiency of stochastically modified bred vectors. In Section~\ref{sec.L96DA} the forecast skill and the reliability of each ensemble type is analysed for ensembles generated from an analysed field, obtained form incorporating imperfect observations in a data assimilation procedure. We conclude in Section~\ref{sec.summary} with a discussion and an outlook.\\

%%%%%%%%%%%%%%%%%%%%%%%%%%%%%%%%%%%%%%%%%%%%%%%%%%%%%%%%%%%%%%%%%%%%%

\section{The Lorenz 96 system}
\label{sec.model}
\begin{figure}[h]
		\centering
		\includegraphics[width=19pc]{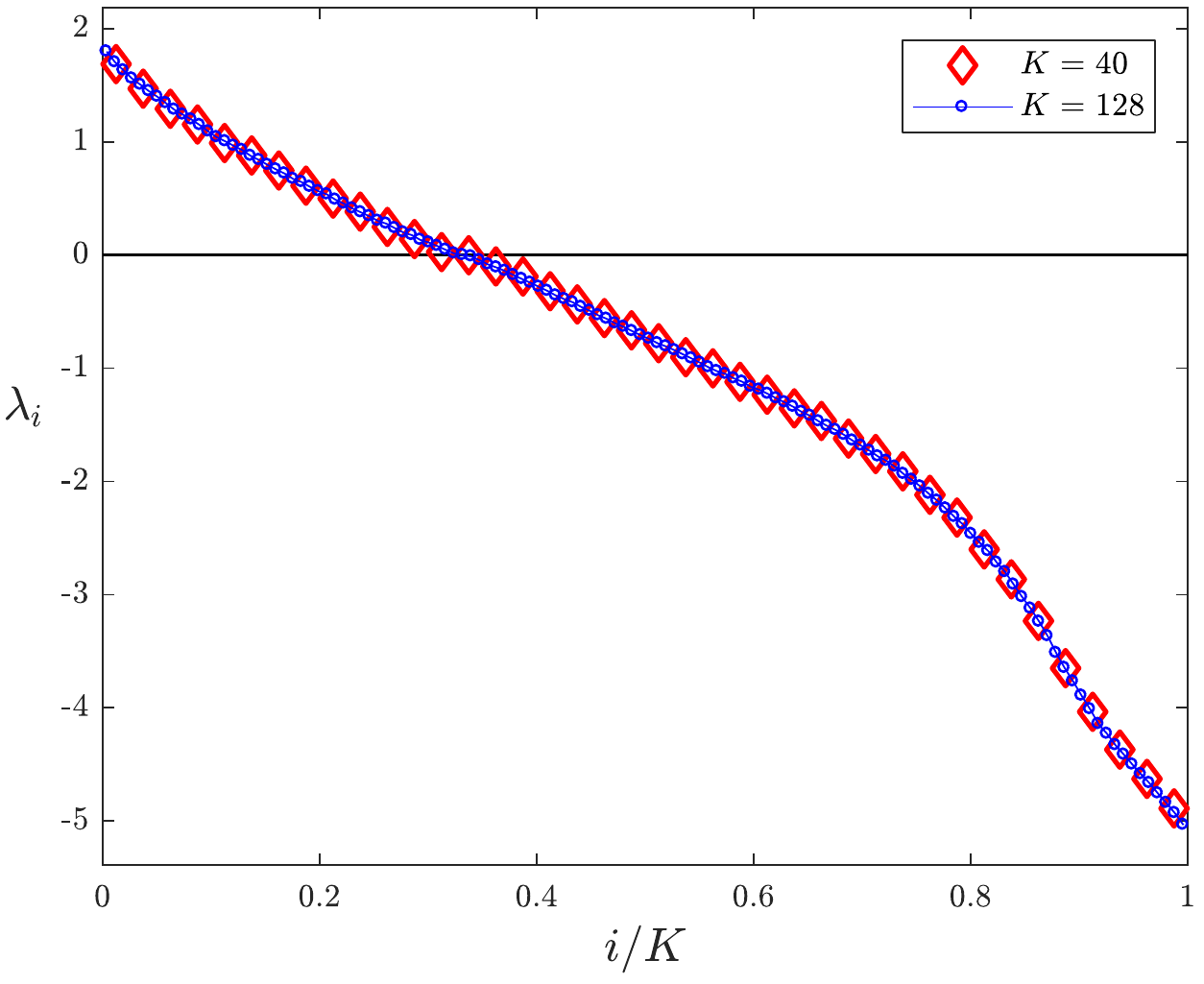}
		\caption{Lyapunov exponent spectrum for the L96 model \eqref{e.L96} with $F=8$ for $K=40$ and $K=128$.}
		\label{fig:L96LyapExp}
	\end{figure}

The Lorenz 96 (L96) system  \citep{Lorenz96} 
\begin{align}
\frac{d}{dt}X_k &= -X_{k-1}(X_{k-2}-X_{k+1})-X_k+F  
\label{e.L96}
\end{align}
with cyclic boundary conditions $X_{k+K}=X_k$ for $k=1,\dots,K$, was introduced as a caricature for the midlatitude atmosphere and has been used  as a test bed for numerous studies in atmospheric sciences. The dynamics of the Lorenz 96 system is characterized by energy conserving nonlinear transport, linear damping and forcing. Despite its simplicity the L96 model exhibits many dynamical scenarios also observed in actual geophysical fluid flows such as regimes and transitions between them \citep{Lorenz06}. The variables $X_k$ can be interpreted as large scale atmospheric fields arranged in the midlatitudes on a latitudinal circle of $30,000$${\rm{km}}$, such as synoptic weather systems. The classical choice $K=40$ corresponds to a spacing between adjacent variables of roughly the Rossby radius of deformation of $750$\,${\rm{km}}$. We shall also consider $K=128$ which implies a spacing between adjacent sites of $234$\,${\rm{km}}$. In both cases we use as forcing amplitude $F=8$ which implies chaotic dynamics \citep{Lorenz98}. The setting with $K=40$ reproduces dynamical patterns with a realistic number of Rossby-like waves and is frequently used in the context of data assimilation. On the other hand, the choice $K=128$ is used to study intrinsic properties of spatially extended dynamical systems \citep{PazoEtAl13}. For $F=8$ the climatic variance is estimated as $\sigma^2=13.25$ and the decorrelation ($e$-folding) time is $\tau=0.41$, for both $K=40$ and $K=128$. The maximal Lyapunov exponent is measured as $\lambda_{\rm{max}}=1.69$ for $K=40$ and as $\lambda_{\rm{max}}=1.775$ for $K=128$. The L96 system is extensive \citep{Karimi10}, in the sense that many relevant quantities (such as surface width, attractor dimension, entropy) scale linearly with the size $K$ and the Lyapunov exponents converge to a continuous function in the limit $K\to\infty$. This is illustrated in Figure~\ref{fig:L96LyapExp}, where we show the Lyapunov spectrum for $K=40$ and $K=128$. For $K=40$ there are $13$ distinct positive Lyapunov exponents, while for $K=128$ there are $42 \approx \frac{128}{40}\times 13$ distinct positive Lyapunov exponents.\\

To numerically simulate the L96 system we employ a fourth-order Runge-Kutta method with a fixed time step $dt = 0.005$. In our simulations an initial transient time of $5000$ time units is discarded to assure that the dynamics has settled on the attractor. 	

%%%%%%%%%%%%%%%%%%%%%%%%%%%%%%%%%%%%%%%%%%%%%%%%%%%%%%%%%%%%%%%%%%%%%

\section{Bred vectors and the breeding method}
\label{sec.BV}
We briefly review the classical breeding method introduced by \cite{TothKalnay93,TothKalnay97}. We closely follow the exposition from our previous work \citep{GigginsGottwald19}. BVs are finite-size, periodically rescaled perturbations generated using the full non-linear dynamics of the system. Centred around a control trajectory $\boldsymbol{z}_c(t_i)$ at some time $t_i$, perturbed initial conditions of size $\delta$, 
\begin{align*}
\boldsymbol{z}_p(t_i) = \boldsymbol{z}_c(t_i) + \delta \frac{\boldsymbol{p}}{\| \boldsymbol{p} \|},
%\label{BVdef1}
\end{align*}
are defined where $\boldsymbol{p}$ is an initial arbitrary random perturbation. The control and the perturbed initial conditions are simultaneously evolved using the full non-linear dynamics for an integration time $T$ until time $t_{i+1}=t_i+T$. At the end of the integration window the difference between the control and the perturbed trajectories 
\begin{align*}
\Delta \boldsymbol{z}(t_{i+1}) = \boldsymbol{z}_p(t_{i+1}) - \boldsymbol{z}_c(t_{i+1})
\end{align*} 
is determined, and the bred vector is defined as the difference rescaled to size $\delta$ with
\begin{align*}
\boldsymbol{b}(t_{i+1}) = \delta \frac{\Delta \boldsymbol{z}(t_{i+1})}{\| \Delta \boldsymbol{z}(t_{i+1}) \|}.
\end{align*}
The perturbation $\boldsymbol{b}(t_{i+1})$ then determines the initial condition of the perturbed trajectory $\boldsymbol{z}_p(t_{i+1}) = \boldsymbol{z}_c(t_{i+1}) + \boldsymbol{b}(t_{i+1})$ at the start of the next breeding cycle. This process of breeding is repeated for several cycles until the perturbation maintains a sufficiently large growth rate and until the perturbations converge in the sense that at time $t_n$ an ensemble of BVs spans the same subspace as BVs obtained in a breeding cycle which had been initialised further in the past. The characteristic time scales of the instabilities of interest and length of the breeding cycles determine how many breeding cycles are required to achieve convergence \citep{TothKalnay93}. For the L96 system a breeding cycle length of $T=0.05$ time units is employed for all simulations, and we employ a spin-up time for the BVs of $500$ time units (which amounts to $10000$ breeding cycles). An ensemble of $N$ BVs is created by $N$ independent breeding cycles initialised from independent initial perturbations $\boldsymbol{p}$. The resulting converged BV ensemble at time $t_i$ is then employed as initial conditions for ensemble forecasts. The breeding method is conceptually similar to the method for generating Lyapunov vectors. They differ though in that Lyapunov vectors are generated using the linearised dynamics and an infinitesimal perturbation $\delta$, whereas BVs are generated using the full nonlinear model and finite perturbation sizes. In contrast to covariant Lyapunov vectors which are mapped by the linear tangent dynamics onto each other, the dynamics of finite-size BVs is not given by a linear mapping and as such they technically do not form a vector space. Despite the similarities between BVs and Lyapunov vectors, we adopt here the point of view outlined in \cite{GigginsGottwald19} that for probabilistic ensemble forecasts the object of interest are the perturbed states $\boldsymbol{z}_p = \boldsymbol{z}_c + \Delta \boldsymbol{z}$, which constitute the sample points for the Monte-Carlo approximation of the probability density function, rather than the differences $\Delta \boldsymbol{z}$.\\ 
%We nevertheless follow the general convention and label them as vectors. \\%We shall investigate in how far BVs are close to Lyapunov vectors in Section~3\ref{sec.CLV}.\\

\begin{figure}[h]
	\centering
	\includegraphics[width=19pc]{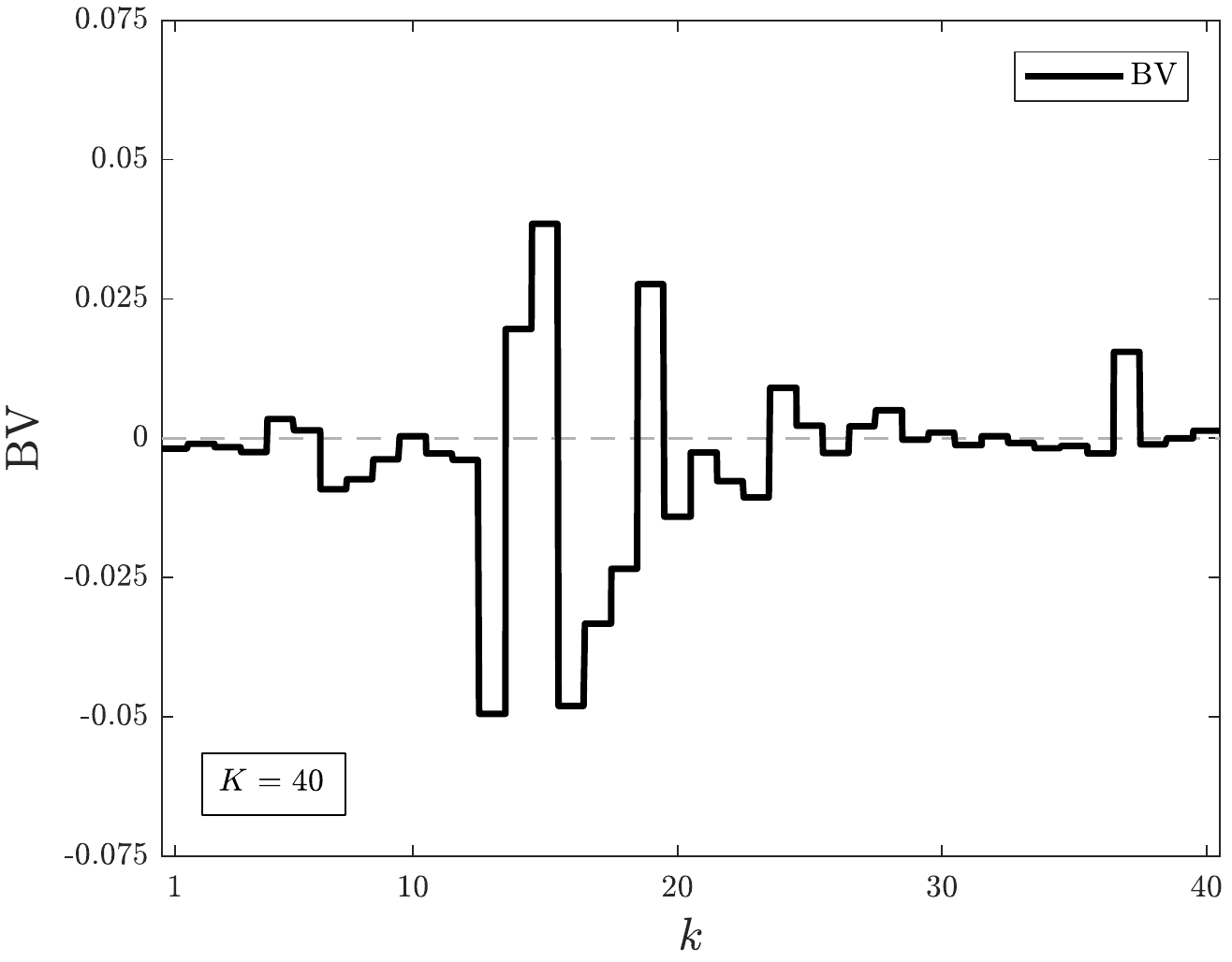}\\
	\vspace{1mm}
	\includegraphics[width=19pc]{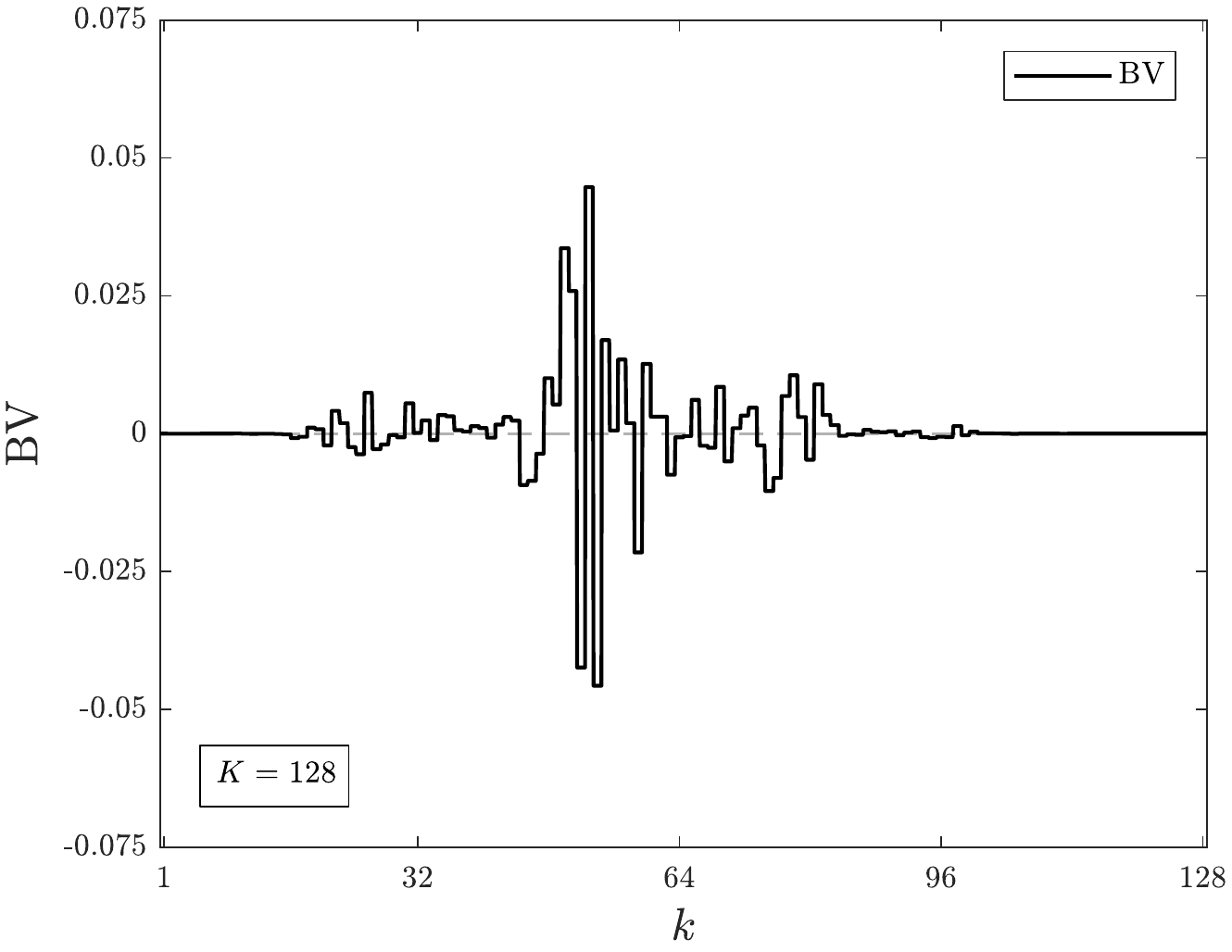}
	\caption{Bred vectors of the L96 system with perturbation size $\delta=0.1$ for different system size $K$. Top: $K=40$. Bottom: $K=128$.}
	\label{fig:L96BVSnapshot}
\end{figure}

\begin{figure}
	\centering
	\includegraphics[width=19pc]{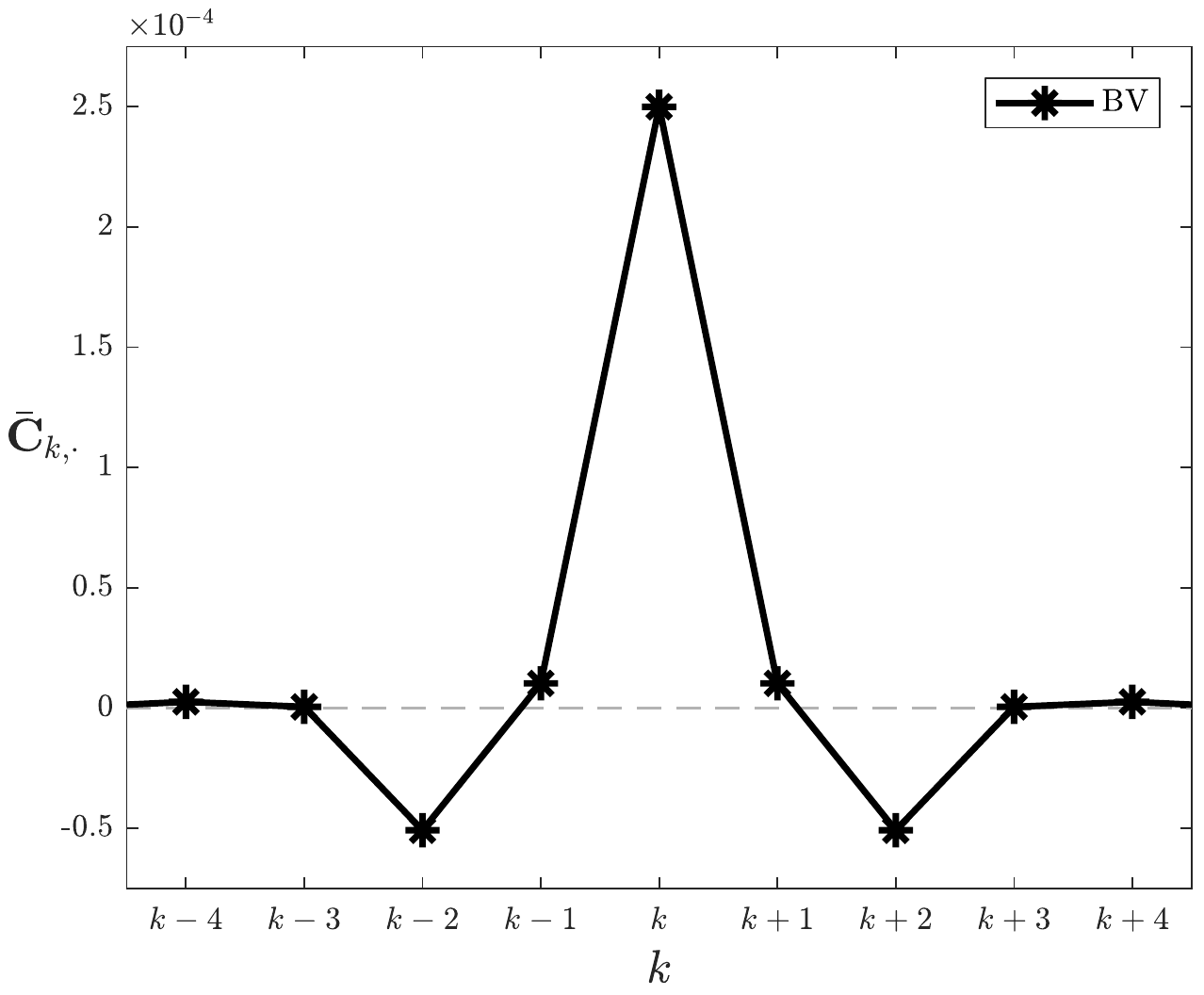}
	\caption{Row-averaged $\bar{\mathbf{C}}_{k,\cdot}$ of the covariance matrix \eqref{e.BVCov} for the  BV of the L96 system with perturbation size $\delta=0.1$ displayed in Figure~\ref{fig:L96BVSnapshot} for $K=40$. For $K=128$ the plot is indistinguishable by eye.}
	\label{fig:L96K40BVCov}
\end{figure}

The bred vectors of the L96 system for system sizes $K=40$ and $K=128$ have markedly different spatial structures. This is illustrated in Figure~\ref{fig:L96BVSnapshot} where snapshots of a typical BV are shown for $K=40$ and for $K=128$ for $\delta=0.1\approx 0.275\, \sigma_{\rm{clim}}$. For the larger $K=128$ system BVs are strongly localised with only a well-defined group of sites having significant entries, whereas for $K=40$ the localisation is less well defined and the size of the active sites with significant entries almost spans the whole domain. Note that the number of "active" sites with increased absolute value of the BVs are roughly the same in both cases, reflecting that BVs capture the same instability, which exhibits the same spatial organization in both cases, but with a better resolution in the larger domain. As we will see, the degree of localisation plays a crucial role for the performance of stochastically perturbed bred vectors. 
To measure the spatial organisation of BVs we consider the $K \times K$ covariance matrix
\begin{align}
\mathbf{C} = \frac{\boldsymbol{b}(t) \big[\boldsymbol{b}(t)\big]^{\mathsf{T}}}{\| \boldsymbol{b}(t) \|_2 \| \boldsymbol{b}(t) \|_2} , 
\label{e.BVCov}
\end{align}
and determine its average $\bar{\mathbf{C}}= \langle \mathbf{C} \rangle$, where the average is taken over realisations of independent BVs generated at different points in time. Since all components for the L96 system are statistically equivalent the $k$-th and $(k+l)$-th rows of $\bar{ \mathbf{C}}$ are identical up to a shift of $l$ components. Figure~\ref{fig:L96K40BVCov} shows the row-averaged ${\bar{\mathbf{C}}}_{k,\cdot}$ of the matrix $\bar{\mathbf{C}}$ for $K=40$ for some arbitrary component $k$ for the BV depicted in Figure~\ref{fig:L96BVSnapshot} (for $K=128$ the correlation structure is identical).\\

In probabilistic forecasting the aim is to approximate the density $\rho(X,\tau)$ at lead time $\tau$ given an initial density $\rho(X,t=0)$, describing the current estimate of the system. Adopting our point of view that BVs are designed to represent a good Monte Carlo estimate of $\rho(X,t=0)$, the property of BVs to capture fast growing dynamically relevant instabilities is translated into the initial conditions associated with BVs which are then likely to be observed at later times $\tau$ representative of the density $\rho(X,\tau)$. The capability of BVs to form an ensemble of independent initial conditions suited for a reliable probabilistic ensemble forecast, depends crucially on the perturbation size $\delta$. For too large perturbation sizes $\delta$, the initial conditions resemble random draws from the attractor (after a typically rapid transition towards it) and the forecast skill deteriorates. Contrary, for too small values of the perturbation size, BVs align with the leading Lyapunov vector (LLV) exhibiting ensemble collapse to a single ensemble member. An ensemble of $N=20$ BVs with $\delta=0.1$, which were initialised with different random perturbations, collapses and the ensemble members are indistinguishable by eye from the ones depicted in Figure~\ref{fig:L96BVSnapshot} for both $K=40$ and $K=128$. Such a lack of diversity of an ensemble of bred vectors presents a major draw back of bred vectors in ensemble forecasting. 
%For example, \cite{WangBishop03} observed the collapse of BVs onto the LLV in the NCAR Community Climate Model. 
In the language of probabilistic forecasts the alignment of BVs with the LLV implies that only a single draw from $\rho(X,t=0)$ is considered. In the next section we present a method how to overcome this drawback while still preserving the desirable features of BVs such as their low computational cost and their dynamical consistency \citep{PazoEtAl10}.

%%%%%%%%%%%%%%%%%%%%%%%%%%%%%%%%%%%%%%%%%%%%%%%%%%%%%%%%%%%%%%%%%%%%%

\section{Stochastically perturbed bred vectors}
\label{sec.SPBV}

\begin{figure}[h]
	\centering
	\includegraphics[width=19pc]{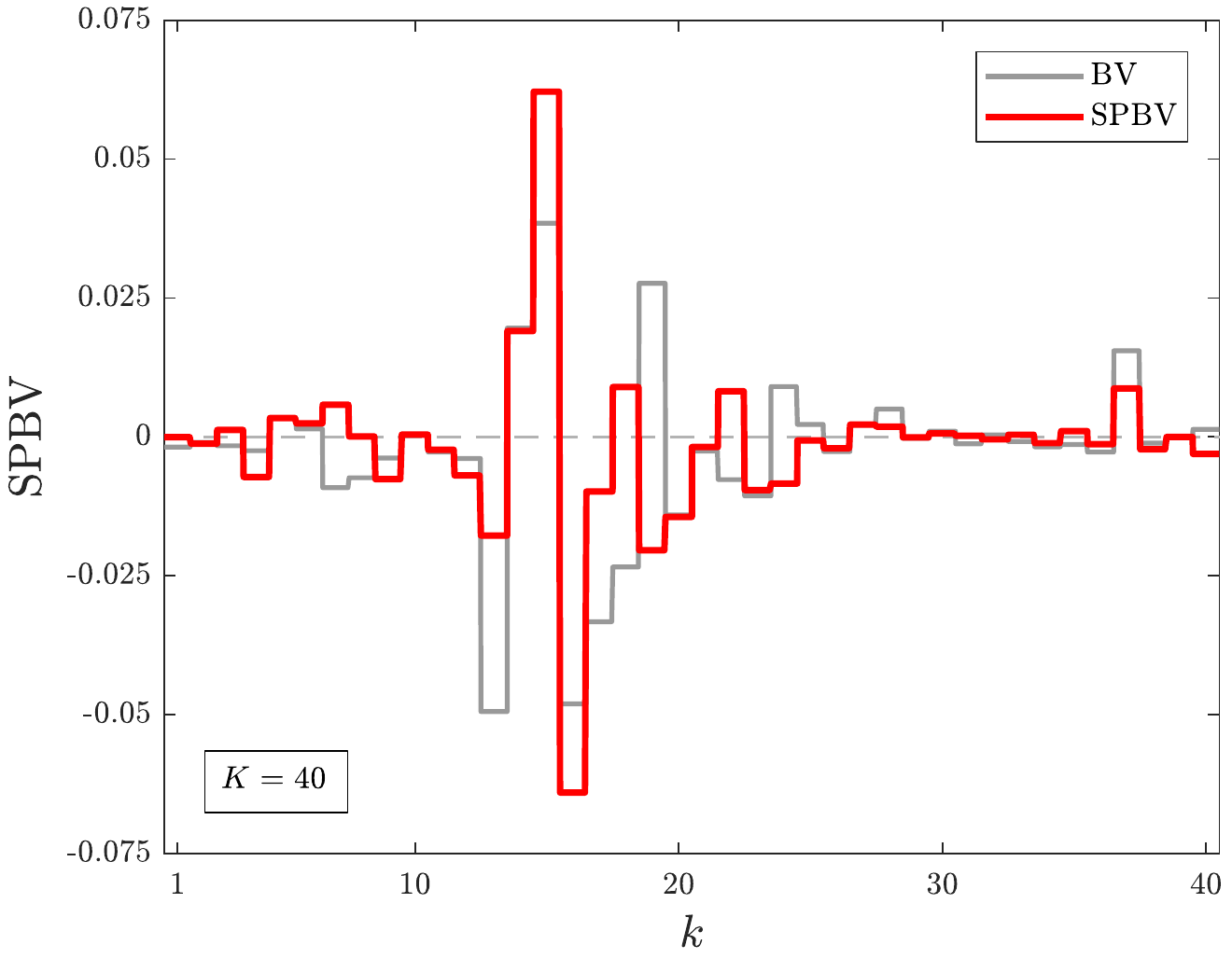}\\
	\vspace{1mm}
	\includegraphics[width=19pc]{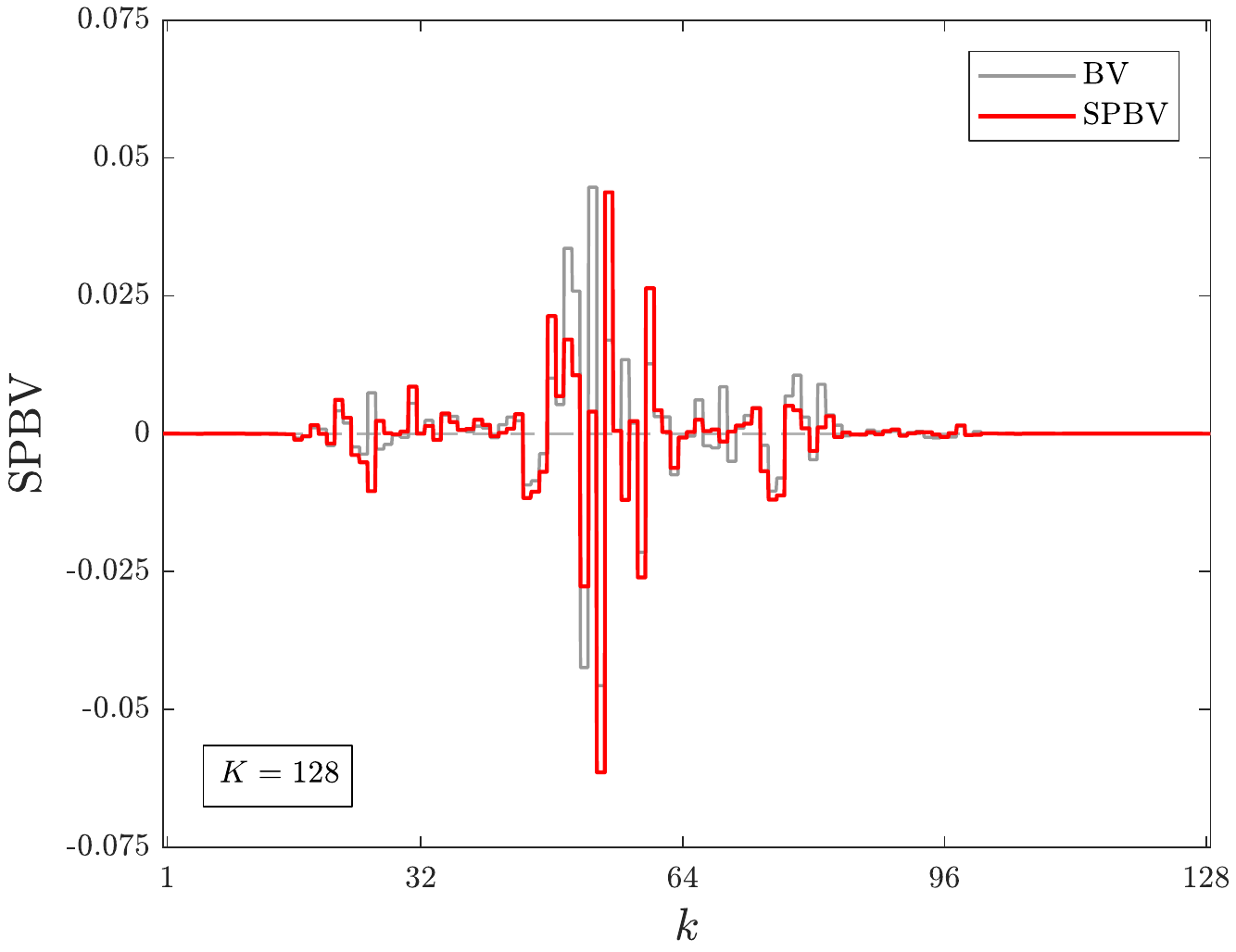}
	\caption{SPBV and its associated parent ${\rm{BV}}$ for the L96 model for perturbation size $\delta=0.1$. Top: $K=40$. Bottom: $K=128$.}
	\label{fig:L96SPBVSnapshot}
\end{figure}

We review here the method proposed in \cite{GigginsGottwald19} to increase the diversity of BV ensembles for multi-scale systems and apply it to systems without scale separation. To generate a diverse ensemble of initial conditions conditioned on the current state of the system, BVs are generated from a parent BV by applying a multiplicative stochastic perturbation to it. The key idea to generating additional draws from the initial density function $\rho(X,t=0)$ is to exploit the fact that in spatially extended dynamical systems, BVs are often localised (as shown in Figure~\ref{fig:L96BVSnapshot} for $K=128$) or exhibit some non-trivial spatial structure (as shown in Figure~\ref{fig:L96K40BVCov} for $K=40$), corresponding to some degree of spatial organisation of error growth. Stochastically perturbed bred vectors (SPBVs) are designed to preserve the spatial structure of BVs, which is paramount to conditioning the initial density $\rho(X,t=0)$ on the current state $X$. SPBVs are defined as
\begin{align}
\boldsymbol{b}_{sp} = \delta \frac{(\mathbf{I}+ \bm{\Xi}) \boldsymbol{b}}{\| (\mathbf{I}+ \bm{\Xi}) \boldsymbol{b} \|}, \label{e.L96SPBVgen}
\end{align}
where $\mathbf{I}$ is the $K \times K$ identity matrix. The diagonal $K \times K$ matrix  $\bm{\Xi}$ with entries $\xi_{jj} \sim \mathcal{N}(0,\sigma^2)$ for $j=1,\ldots,K$ with variance parameter $\sigma^2$ represents the stochastic perturbation. The stochastic perturbation is performed only once as a post-processing step from a given parent BV when generating initial conditions for a forecast ensemble, and therefore does not significantly add to the computational cost. In Figure~\ref{fig:L96SPBVSnapshot} we show a realisation of an SPBV, overlaid with their parent BV, for a perturbation size of $\delta=0.1$ with noise strength $\sigma = 1.25$, for $K=40$ and for $K=128$. It is clearly seen that the spatial structure of the perturbation is preserved.\\   

The stochastic perturbations generate initial conditions that are nearby the attractor and after a typically rapid relaxation towards the attractor along the stable manifold, approach the attractor close in phase space to the initial condition associated with the parent BV, which we know is capturing fast error growth. The stochasticity hence allows to sample the phase space on the attractor in the fastest growing region.\\

\begin{figure}
	\centering
	\includegraphics[width=19pc]{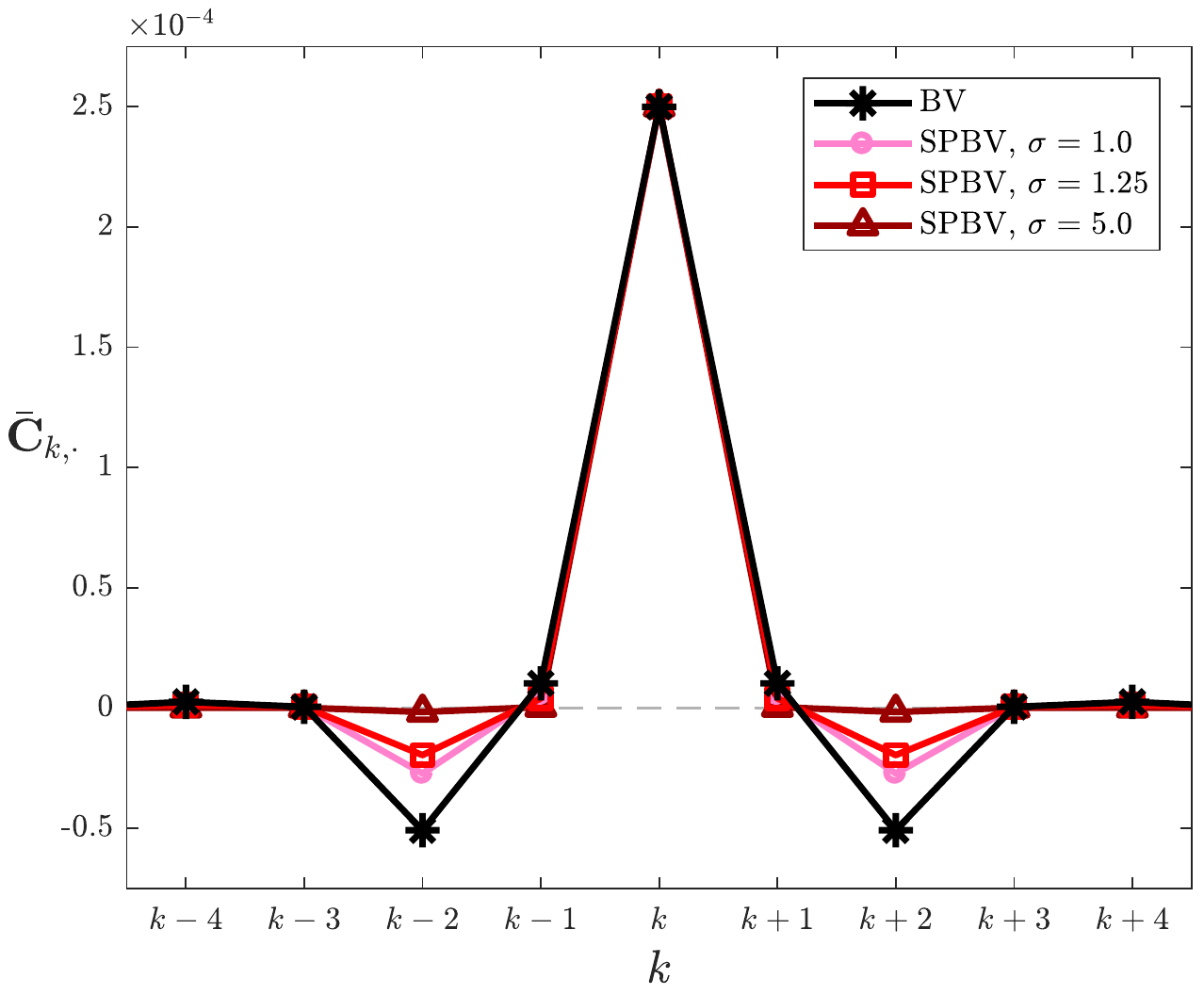}
	\caption{Row-averaged $\bar{\mathbf{C}}_{k,\cdot}$ from the covariance matrix \eqref{e.BVCov} for SPBVs of the L96 system with various noise strengths $\sigma = 1.0$, $\sigma = 1.25$ and $\sigma = 5.0$ for $K=40$. The covariance of BVs is depicted as a reference. For $K=128$ the plot is indistinguishable by eye.}
	\label{fig:L96K40BVCovSigma}
\end{figure}

The noise strength $\sigma$ obviously plays a central role. When $\sigma \to 0$ SPBVs essentially reproduce the parent BVs they were generated from, and the spatial structure is exactly preserved but no diversity is gained. In the other extreme case $\sigma \to \infty$, the behaviour depends on the degree of localisation. For the strongly localised case $K=128$ with many vanishing BV components (cf. Figure~\ref{fig:L96SPBVSnapshot} (bottom)), the degree of localisation remains preserved since SPBVs are rescaled to size $\delta$, and the diversity is greatly enhanced. This is the case discussed in the multi-scale setting in \cite{GigginsGottwald19}. The weakly localised case when there are no significant regions with vanishing components of the BV (cf. Figure~\ref{fig:L96SPBVSnapshot} (top)) is more complex. For sufficiently large magnitudes of the noise strength $\sigma$, SPBVs become spatially uncorrelated random perturbations of size $\delta$. This allows for (almost) maximal diversity of the ensemble which, however, comes at the cost of destroying the inherent spatial structure of the dynamically relevant fast growing perturbations. The destruction of the spatial structure implies that we typically do not sample the phase space region locally but instead generate initial conditions as random draws from the attractor, which are not conditioned on the current state. In Figure~\ref{fig:L96K40BVCovSigma} we illustrate the loss of spatial structure by showing the average of rows $\bar{\mathbf{C}}_{k,\cdot}$ of  the covariance matrix \eqref{e.BVCov} for SPBVs for increasing values of the noise strength $\sigma$. It is seen that for $\sigma = 1.0$ and for $\sigma = 1.25$ the nontrivial correlations between adjacent sites are preserved albeit reduced in magnitude, whereas for $\sigma=5$ the spatial structure is entirely lost and adjacent sites are uncorrelated.\\
% This suggests that $\sigma$ needs to be tuned in the weakly localised case to balance diversity and reliability of the ensemble with forecast skill and root-mean-square errors of the forecast. We shall provide numerical evidence for these statements in Section~\ref{sec.L96tr} after introducing the relevant diagnostics in Section~\ref{sec.diagnostics}.\\

\begin{figure}[h]
	\centering
	\includegraphics[width=19pc]{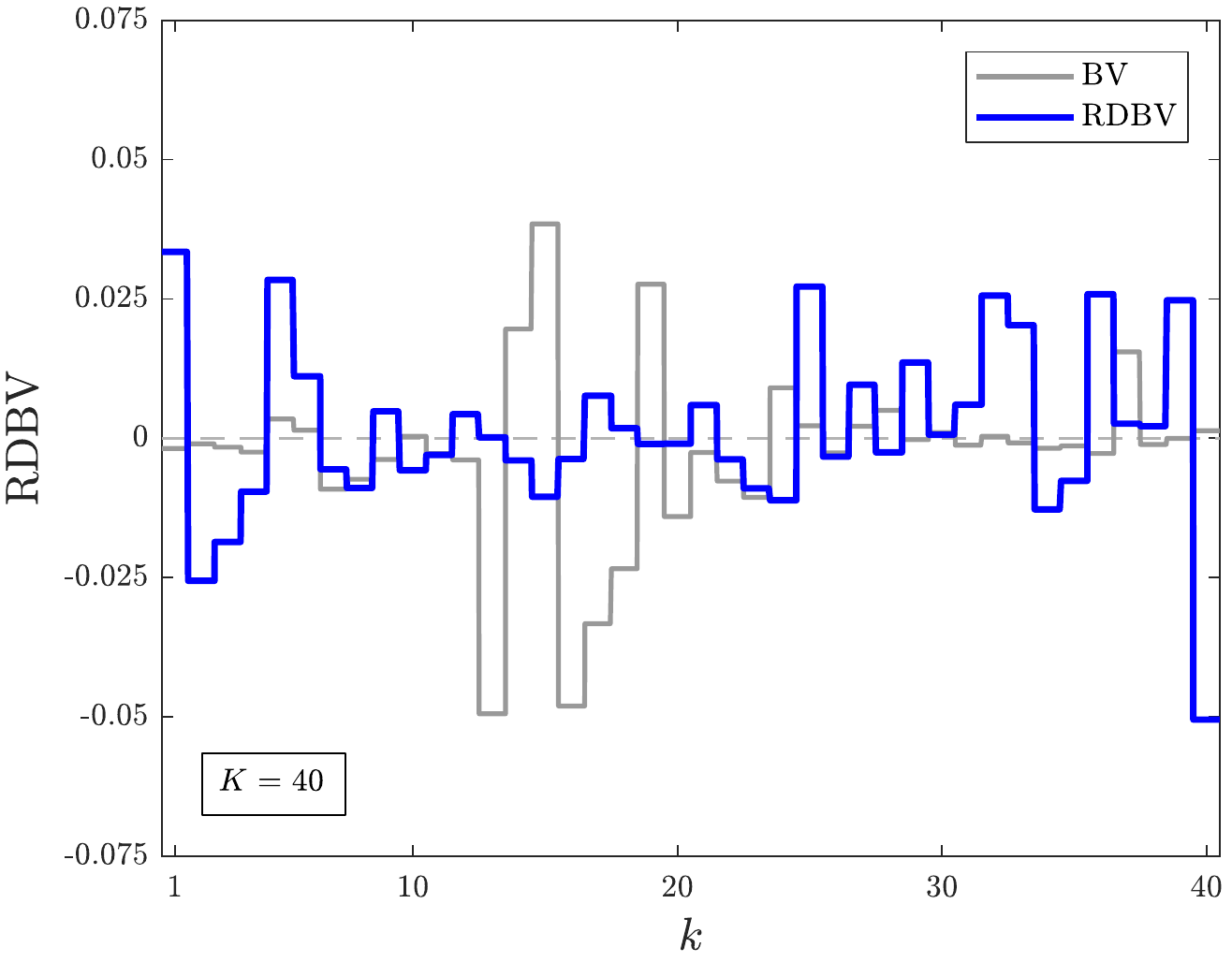}\\
	\vspace{1mm}
	\includegraphics[width=19pc]{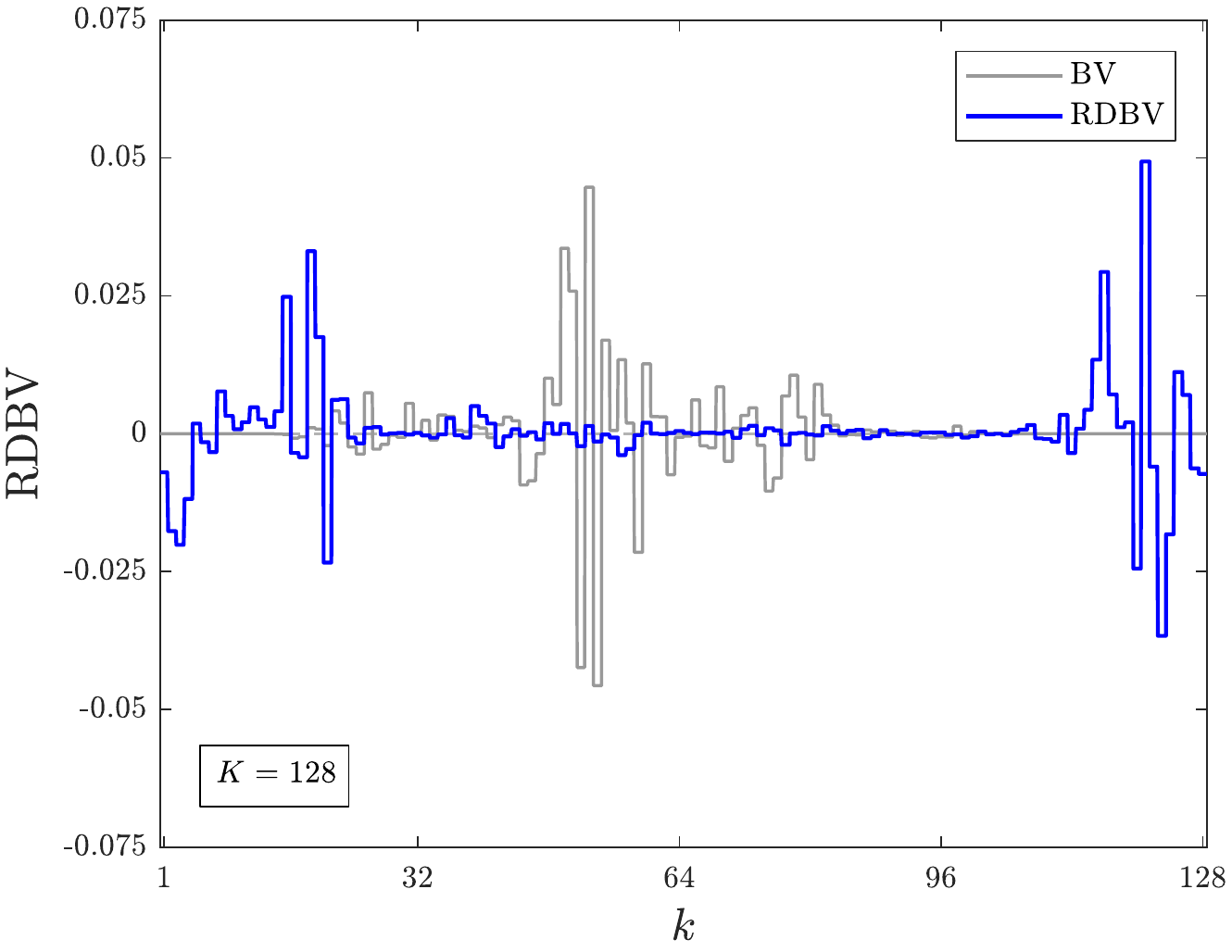}
	\caption{RDBV and BV for the L96 model for perturbation size $\delta=0.1$. The BVs are as in Figure~\ref{fig:L96BVSnapshot}. Top: $K=40$. Bottom: $K=128$.}
	\label{fig:L96RDBVSnapshot}
\end{figure}

We also consider the so called random draw bred vectors (RDBVs) introduced in \cite{GigginsGottwald19}. An ensemble of RDBVs is generated by randomly selecting classical BVs which were generated from independent initial conditions randomly drawn from the attractor. To avoid storing a huge library of independent BVs, an ensemble of RDBVs is generated on the fly by evolving $N$ independent control trajectories started from independent initial conditions, each generating a single BV. Whereas SPBVs are designed to sample the phase space locally, RDBVs are dynamically inconsistent in the sense that they may, after a quick relaxation towards the attractor, evolve into states which are not close in phase space to the current state of the control. Example RDBVs for $K=40$ and $K=128$ are shown in Figure~\ref{fig:L96RDBVSnapshot}. We remark that, contrary to SPBVs, RDBVs form an (almost) orthogonal ensemble.\\ %, regardless of the whether we are in the strongly localised $K=128$ case or the weakly localised $K=40$ case. This is due to the property that each RDBV ensemble member is independently to all other ensemble members, but no orthogonality constraints are enforced. \\

%%%%%%%%%%%%%%%%%%%%%%%%%%%%%%%%%%%%%%%%%%%%%%%%%%%%%%%%%%%%%%%%%%%%%

\subsection{Dynamic properties of bred vectors: Backward and covariant Lyapunov vectors}
\label{sec.CLV}

We now probe how bred vectors and their stochastic modifications relate to dynamically relevant modes such as Lyapunov vectors which capture the asymptotic growth of infinitesimal perturbations, and thereby in how far they are dynamically adapted. The dynamic adaptivity of classical BVs was established in \cite{PazoEtAl10}. We now show that SPBVs inherit this property from their parent BVs. In particular, we consider the relationship between bred vectors and backward Lyapunov vectors and covariant Lyapunov vectors. Backward Lyapunov vectors are initialised in the asymptotically distant past and are generated by solving the linear tangent model of the dynamical system under a Gram-Schmidt orthogonalisation procedure. The orthogonal backward Lyapunov vectors are not covariant under the linear tangent dynamics and all of them typically evolve under the dynamics into the leading Lyapunov vector (LLV). Covariant Lyapunov vectors, on the contrary, form a typically non-orthogonal basis of the tangent space and are mapped onto each other by the linearised tangent dynamics. The associated asymptotic growth rates of backward and covariant Lyapunov vectors, the Lyapunov exponents, are shown in Figure~\ref{fig:L96LyapExp} for the L96 system. As BVs, the first few leading covariant Lyapunov vectors exhibit a localised spatial structure in the L96 system (not shown), with strong localisation for $K=128$ and weak localisation for $K=40$.

We quantify the relationship between the respective BV ensembles and Lyapunov vectors by measuring the average projection of BV ensembles onto backward and onto covariant Lyapunov vectors, and consider the following measure for the degree of projection
\begin{align}
\pi^{n}_i(t) = \left\vert 
\frac{\boldsymbol{b}^{n}(t)}{\| \boldsymbol{b}^{n}(t)\|} \cdot  \frac{\boldsymbol{l}_i(t)}{\| \boldsymbol{l}_{i}(t)\|}
\right\vert ,
\label{e.BVLVproj}
\end{align}
where $\boldsymbol{b}^{n}(t)$ denotes the $n$th bred vector ensemble member at time $t$ and $\boldsymbol{l}_i(t)$ denotes the Lyapunov vector corresponding to the $i$th largest Lyapunov exponent at time $t$. We report here on the average degree of projection $\bar\pi_i$ where we average $\pi^n_i(t)$ over time and over the ensemble members. Note that $\bar \pi_i=1$ corresponds to perfect alignment and $\bar\pi_i=0$ corresponds to (on average) no  alignment.\\

There exist several efficient numerical algorithms to calculate the covariant Lyapunov vectors \citep{WolfeSamelson07,GinelliEtAl07}. We use here the algorithm by \cite{GinelliEtAl07} as described in \cite{KuptsovParlitz12} to numerically calculate covariant Lyapunov vectors. We use a spin-up time of $2,500$ time units to converge to the set of backward Lyapunov vectors evolving forward in time, and a further $2,500$ time units to ensure convergence to the set of expansion coefficients of the covariant Lyapunov vectors that express the covariant Lyapunov vectors in the basis of forward and backward Lyapunov vectors respectively, evolving backward in time. Orthonormalisation of the backward Lyapunov vectors is performed at every time step.\\

\begin{figure}
	\centering
	\includegraphics[width=19pc]{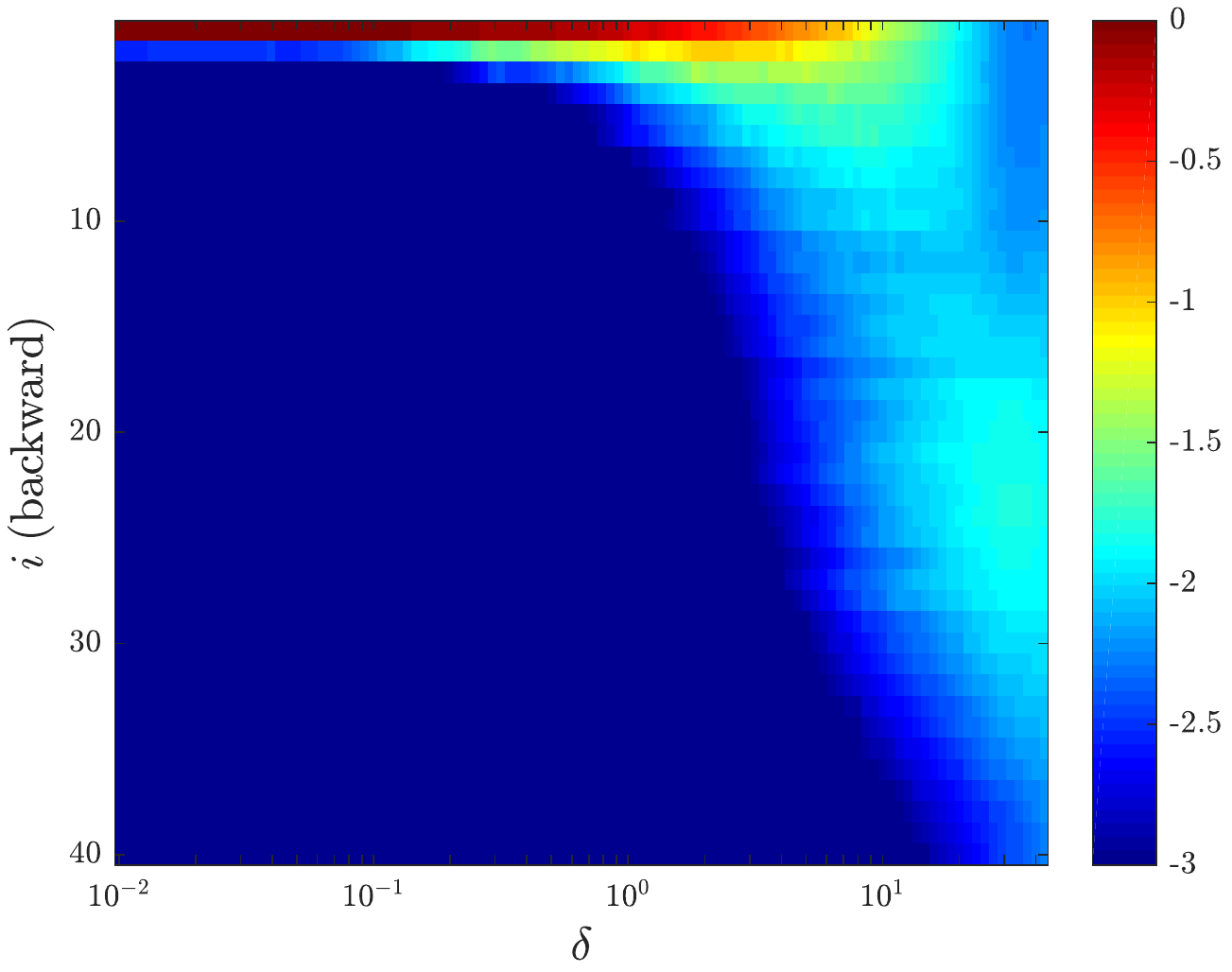}\\
	\vspace{1mm}
	\includegraphics[width=19pc]{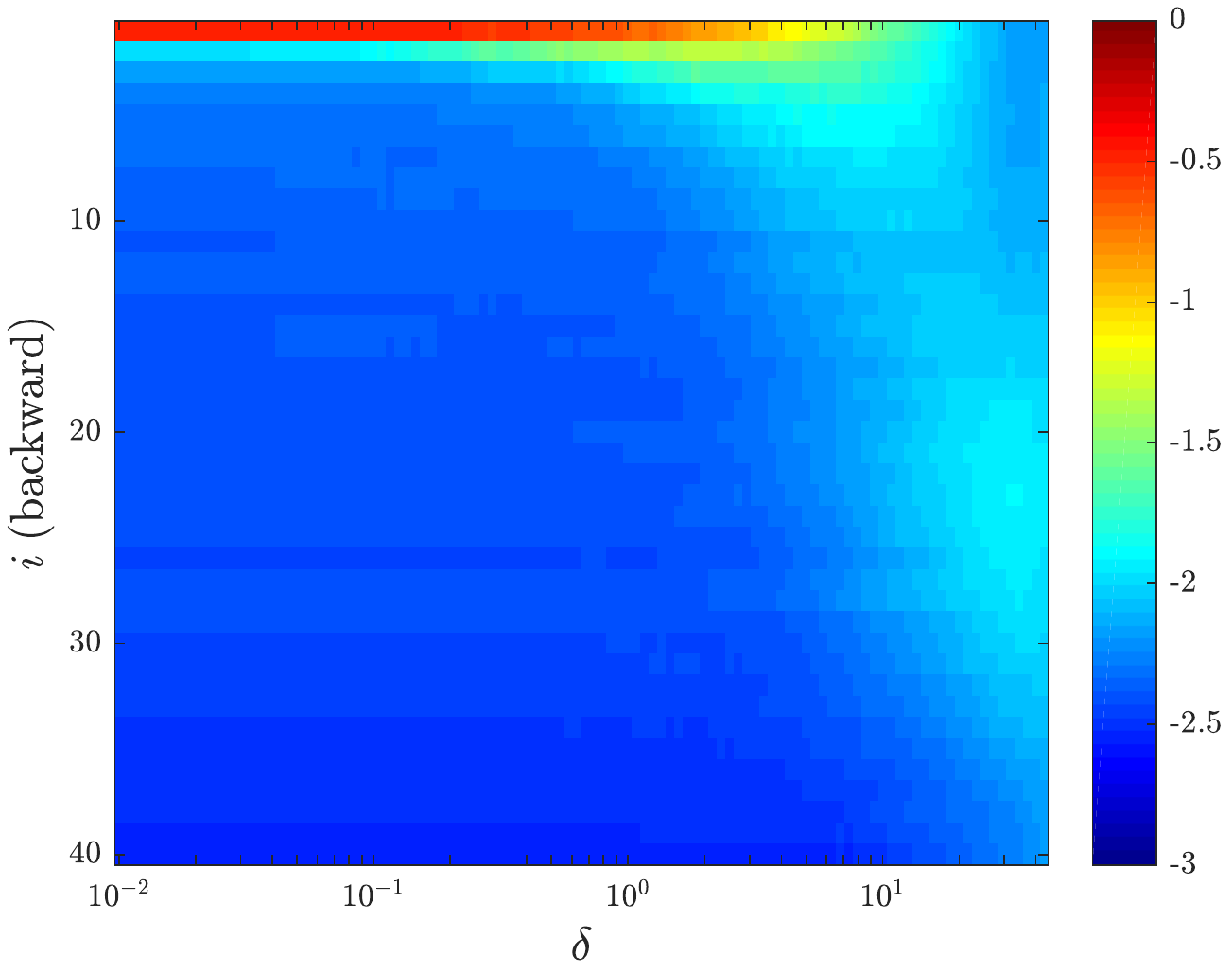}
	\caption{Average projection $\bar\pi_i$ of backward Lyapunov vectors for $K=40$. Results are shown on a logarithmic scale. Top: BVs. Bottom: SPBVs with $\sigma=1.25$.}
	\label{fig:L96K40BLV}
\end{figure}

\begin{figure}
	\centering
	\includegraphics[width=19pc]{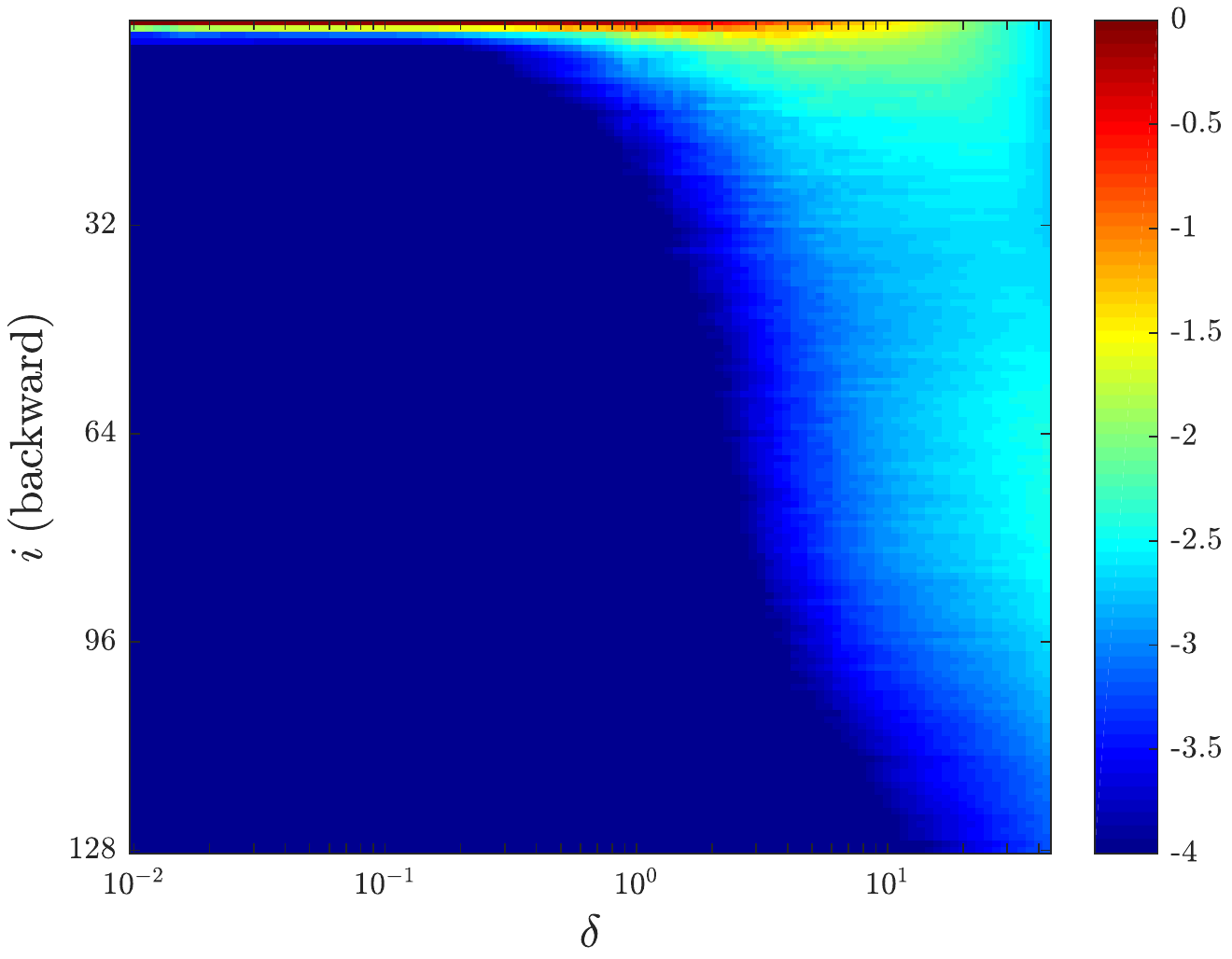}\\
	\vspace{1mm}
	\includegraphics[width=19pc]{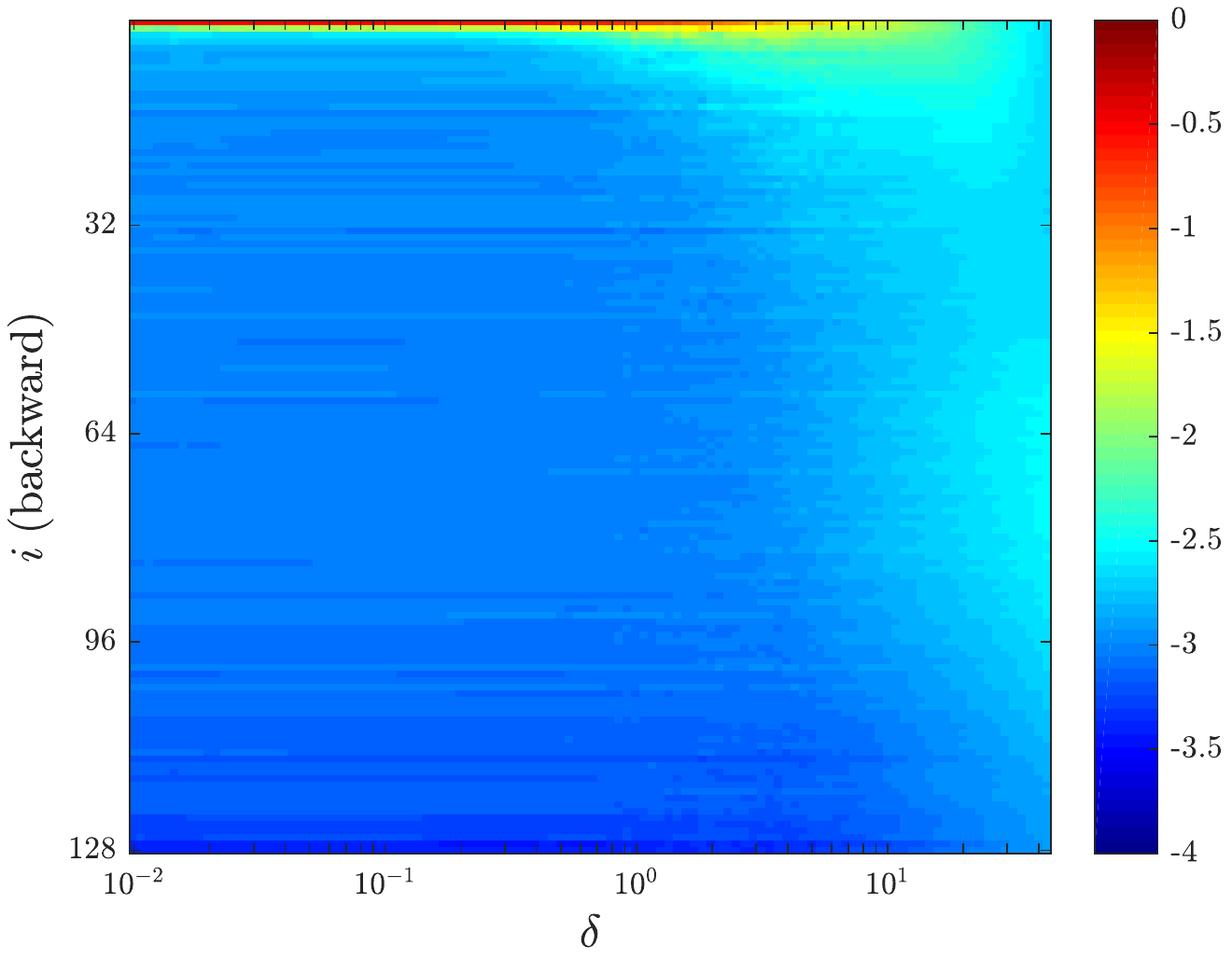}
	\caption{Average projection $\bar\pi_i$ of backward Lyapunov vectors for $K=128$. Results are shown on a logarithmic scale. Top: BVs. Bottom: SPBVs with $\sigma=1.25$.}
	\label{fig:L96K128BLV}
\end{figure}

Figures~\ref{fig:L96K40BLV} and \ref{fig:L96K128BLV} display $\bar{\pi}_i$ for all $i=1,\dots,K$ backward Lyapunov vectors for $K=40$ and $K=128$ respectively, for classical BVs (top) and SPBVs (bottom). It is clearly seen that both, classical BVs and SPBVs, project almost completely onto the first backward Lyapunov vector (the LLV) for small $\delta < 0.1$ and are orthogonal to all other directions for both dimension sizes. When the perturbation size lies between $0.1\lesssim \delta \lesssim 8$ for $K=40$ and between $0.1\lesssim \delta \lesssim 5$ for $K=128$, BVs also project onto the next few backwards Lyapunov vectors. We shall see in Section~\ref{sec.L96DA}, that for these perturbation sizes, BV ensembles have collapsed to a single member and have an ensemble dimension (to be defined below in (\ref{e.ensDimFormula})) strictly equal to $1$ (cf. Figure~\ref{fig:L96EnsDim}). For $\delta >1$ the non-vanishing projections of BVs onto the next Lyapunov vectors stem from increasing fluctuations of the BV ensemble around the LLV. We observe that $\pi^{n}_i(t)$ may strongly fluctuate in time and individual members of a BV/SPBV ensemble may exhibit, locally in time, strong projections on higher Lyapunov vectors. In such cases when BVs/SPBVs do not fully align with the LLV, they lie typically in the subspace spanned by the first few Lyapunov vectors (not shown). For even larger values of the perturbation size, BVs do not significantly project onto the linear Lyapunov vectors as they evolved into truly nonlinear objects. 
%However, there is a small projection onto the Lyapunov vectors centred around modes $22$ for $K=40$ and mode $72$ for $K=128$. 
%We do note that there are some very small projections for extremely large values of $\delta$ corresponding to slowly decaying directions between $20 \le i \le 30$ for $K=40$ and $60 \le i \le 90$ for $K=128$.\\
%%
%%
%\gaginline{For $K=40$ the $\lambda=0$ mode is mode $14$, so our former argument employing the translational symmetry and westward propagation cannot be employed?} 
%%
%%
%SPBV ensembles exhibit similar behaviour. 
% also project onto the backward Lyapunov vectors but naturally feature weaker projections due to the stochastic perturbations. SPBVs tend to project most onto the LLV and significantly weaker projections onto the next few leading backward Lyapunov vectors. 

\begin{figure}
	\centering
	\includegraphics[width=19pc]{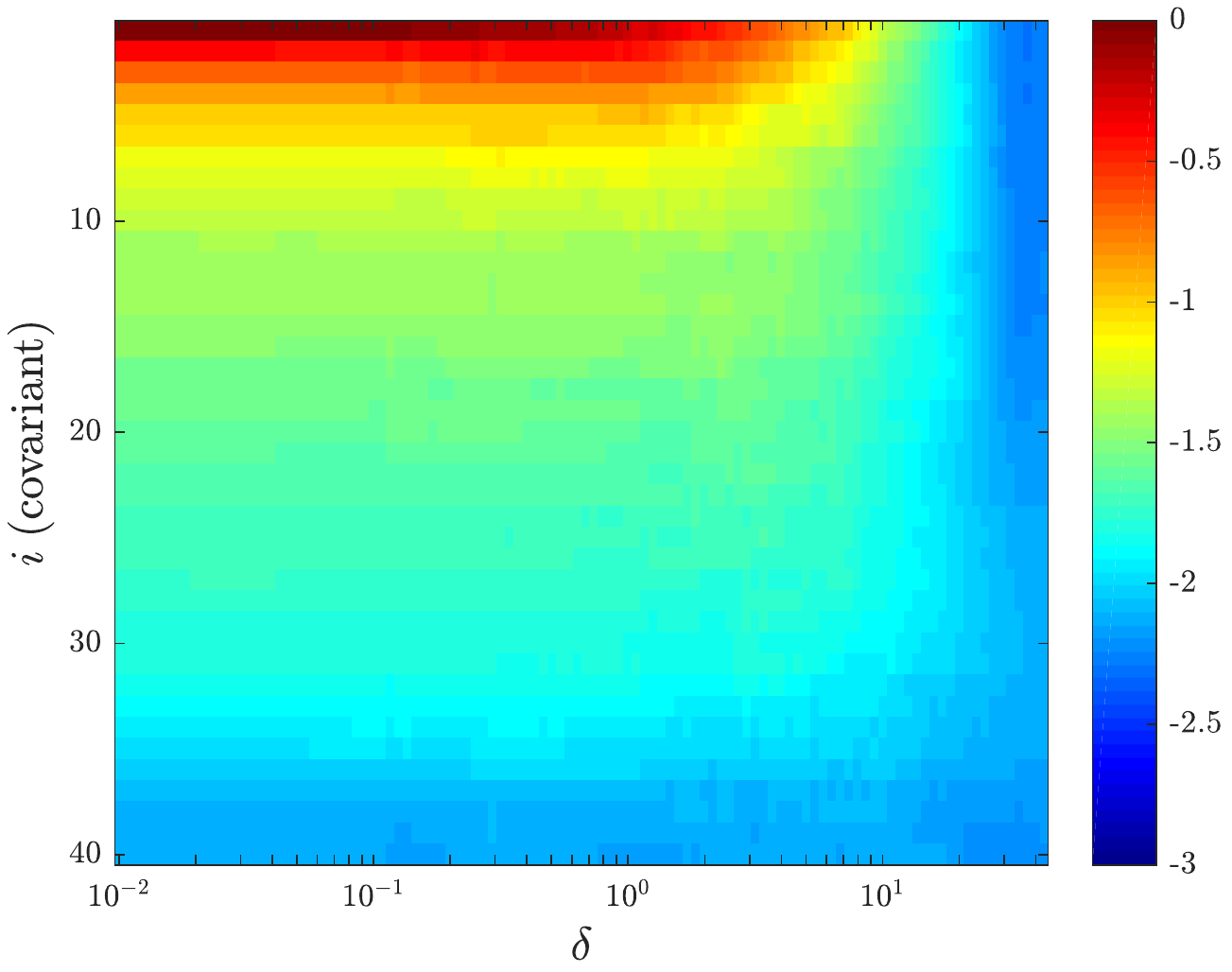}\\
	\vspace{1mm}
	\includegraphics[width=19pc]{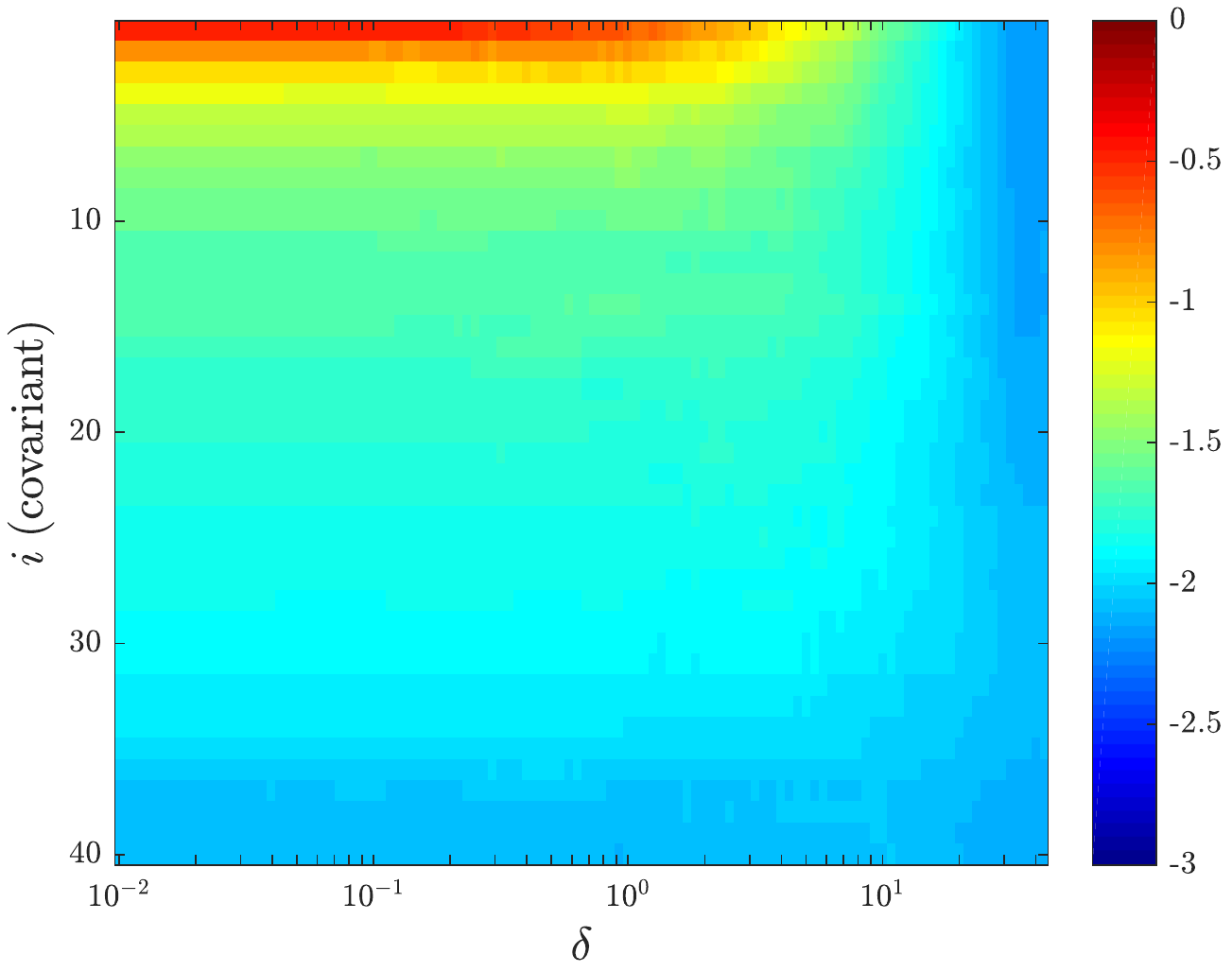}
	\caption{Averaged absolute value projection $\bar\pi_i$ of covariant Lyapunov vectors for $K=40$. Results are shown on a logarithmic scale. Top: BVs. Bottom: SPBVs with $\sigma=1.25$.}
	\label{fig:L96K40CLV}
\end{figure}

\begin{figure}
	\centering
	\includegraphics[width=19pc]{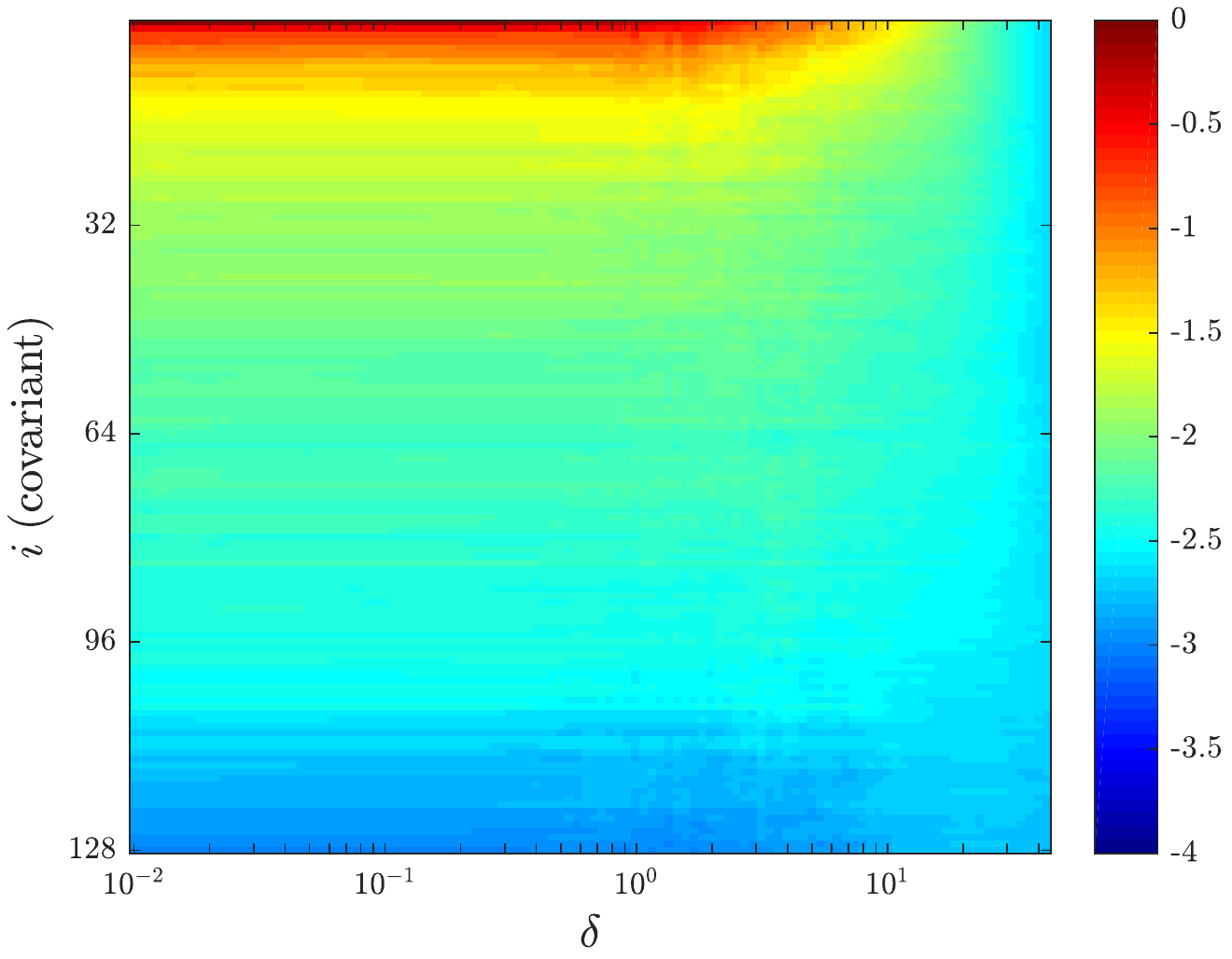}\\
	\vspace{1mm}
	\includegraphics[width=19pc]{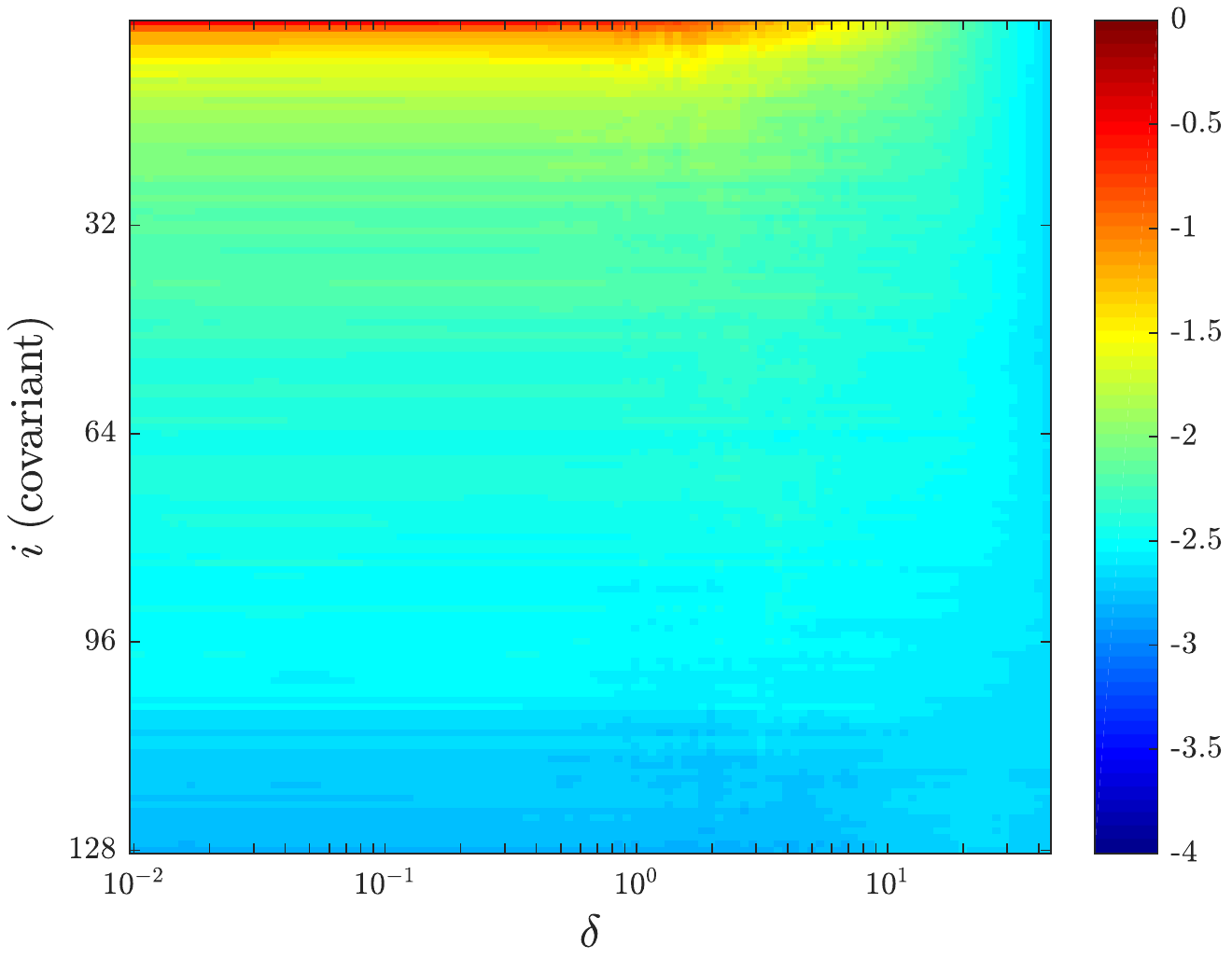}
	\caption{Averaged absolute value projection $\bar\pi_i$ of covariant Lyapunov vectors for $K=128$. Results are shown on a logarithmic scale. Top: BVs. Bottom: SPBVs with $\sigma=1.25$.}
	\label{fig:L96K128CLV}
\end{figure}

Projections of BVs and SPBVs onto covariant Lyapunov vectors exhibit similar signatures as for backward Lyapunov vectors. Contrary to backward Lyapunov vectors, covariant Lyapunov vectors do not form an orthogonal basis. Furthermore, successive covariant Lyapunov vectors are likely to be localised in similar spatial regions to each other whereas this is not the case for backward Lyapunov vectors due to non-dynamical orthogonality constraint \citep{HerreraEtAl11}. Hence fluctuations of BVs cause them to project onto several of the covariant Lyapunov vectors. In particular, we see strong projections of classical BVs onto covariant Lyapunov vectors with index $i \le 7$ for $K=40$, and onto those with index  $i \le 12$ for $K=128$. As for backward Lyapunov exponents, the projection onto dynamically relevant low-index Lyapunov vectors drops off when BVs gain diversity at $\delta \approx 8$ for $K=40$ and at $\delta \approx 5$ for $K=128$. SPBVs feature weaker projections onto the higher-index covariant Lyapunov vectors with significant projections in the smaller range $i \le 3$ and $i \le 6$ for $K=40$ and $K=128$, respectively. This overall stronger projection of SPBVs to the low-index covariant Lyapunov vectors is caused by the ensemble averaging (SPBVs are generated from a single collapsed BV). As for backward Lyapunov vectors, the lower dimensional subspaces onto which BVs and SPBVs project onto can fluctuate over time, in particular for larger values of the perturbation size $\delta$ (not shown). It is pertinent to notice that the projection of SPBVs onto the Lyapunov vectors weakens as $\sigma$ is increased and saturates for sufficiently large $\sigma$, in particular in the weakly localised case $K=40$ (not shown). We shall see below that there is a trade-off between providing the most ensemble diversity while preserving dynamic adaptivity for SPBVs.\\ 

We conclude that BVs and SPBVs share similar localisation structure to that of the first few covariant Lyapunov vectors, and that for sufficiently small noise strength $\sigma$ SPBVs inherit from BVs the desirable property of dynamical adaptivity. RDBVs, on the other hand, do not exhibit any significant average  projections onto any of the backward or covariant Lyapunov vectors as they are unrelated to the local dynamics (not shown). RDBVs are hence dynamically not adapted. We mention that ETKF ensembles are also dynamically adapted. Contrary to BVs and SPBVs which project dominantly onto the first few Lyapunov vectors, ETKF ensembles project homogeneously onto the whole unstable subspace (not shown; see also \citep{NgEtAl11}).

%BVs and SPBVs share similar localisation structure to that of the first few covariant Lyapunov vectors (not shown), and for sufficiently small noise strength $\sigma$ SPBVs inherit from BVs the desirable property of dynamical adaptivity in the sense that their spatial and temporal evolution resembles the actual evolution of error growth of the underlying dynamical system. RDBVs, on the other hand, do not exhibit any significant average  projections onto any of the backward or covariant Lyapunov vectors as they are unrelated to the local dynamics (not shown). RDBVs are hence dynamically not adapted.

%%%%%%%%%%%%%%%%%%%%%%%%%%%%%%%%%%%%%%%%%%%%%%%%%%%%%%%%%%%%%%%%%%%%%	

\section{Diagnostics}
\label{sec.diagnostics}

To illustrate how SPBVs can be used as a reliable diverse ensemble with improved forecast skill we now introduce several diagnostics. In particular, we consider the ensemble dimension to measure the diversity of an ensemble, the root-mean-square error to quantify the forecast skill and several reliability measures to probe the probabilistic properties of an ensemble. This set of diagnostics has previously been used to study the performance of SPBVs and RDBVs \citep{GigginsGottwald19}.

%%%%%%%%%%%%%%%%%%%%%%%%%%%%%%%%%%%%%%%%%%%%%%%%%%%%%%%%%%%%%%%%%%%%%	

\subsection{Ensemble dimension}
\label{sec.Dens}

We quantify the diversity of an ensemble using the "ensemble dimension" \citep{BrethertonEtAl99,OczkowskiEtAl05}, also known as the "bred vector dimension" \citep{PatilEtAl01}. The ensemble dimension is a measure for the dimension of the subspace spanned by a set of vectors. For an ensemble of $N$ BVs $\{\boldsymbol{b}^{(n)}(t)\}_{n=1,\dots ,N}$ at a given time $t$, the ensemble dimension is defined as
\begin{align} 
\mathcal{D}_{ens}(t) = \frac{\Big( \sum_{n=1}^{N}\sqrt{\mu_n} \Big)^2}{\sum_{n=1}^{N}\mu_n}, 
\label{e.ensDimFormula}
\end{align} 
where the $\mu_n$ are the eigenvalues of the $N \times N$ covariance matrix $\mathbf{C}$ (cf. (\ref{e.BVCov})). 
%\begin{align} 
%\mathbf{C}_{nm}(t) = \frac{[\boldsymbol{b}^{(n)}(t)]^{\mathsf{T}}  \boldsymbol{b}^{(m)}(t)}{\| \boldsymbol{b}^{(n)}(t) \|_2 \| %\boldsymbol{b}^{(m)}(t) \|_2}.
%\end{align}
The ensemble dimension takes values between $\mathcal{D}_{ens} = 1$ and $\mathcal{D}_{ens} = {\rm{min}}(N,D)$, where $D$ is the total dimension of the dynamical system, depending on whether the ensemble members are all aligned or are orthogonal to each other. We consider in our numerical experiments the temporal average ${\bar{\mathcal{D}}}_{ens}$ to characterise the diversity of an ensemble. 
%For computational ease, we compute the temporal average as an ensemble average over $M=2500$ independent BV realisations; we have tested that the temporal and ensemble average yield the same results.

%%%%%%%%%%%%%%%%%%%%%%%%%%%%%%%%%%%%%%%%%%%%%%%%%%%%%%%%%%%%%%%%%%%%%	

\subsection{Ensemble forecast skill}
\label{sec.skill}

To measure the forecast skill of an ensemble with $N$ members $X^{(n)}_{k}$, $n=1,\dots ,N$, we evaluate the root-mean-square-error (RMS error) between the truth and the ensemble average. Denoting the ensemble average with angular brackets, we introduce the ensemble mean \begin{align}
\langle X_k \rangle = \frac{1}{N}\sum_{n=1}^N X^{(n)}_{k} 
\end{align}
and the site-averaged root-mean-square error between the truth $X^{tr}_k$ and the ensemble average over $M$ realizations as a function of the lead time $\tau$, 
\begin{align}
\mathcal{E}(\tau) = \sqrt{\frac{1}{M}\sum_{m=1}^{M}\frac{1}{K}\sum_{k=1}^{K}
	\| X^{tr}_{k,m}(\tau) - \langle{X}_{k,m}\rangle (\tau) \|^2},
	\label{e.RMSErr}
\end{align}
where the index $m=1,\ldots,M$ labels the realisation. Similarly, to quantify the dispersion of the ensemble, we consider the site-averaged root-mean-square spread (RMS spread) 
\begin{align}
\mathcal{S}(\tau) = \sqrt{
\frac{1}{M}\sum_{m=1}^{M}
	\frac{1}{K}\sum_{k=1}^{K}
\langle
	\| X^{(n)}_{k,m}(\tau) - \langle{X}_{k,m}\rangle (\tau) \|^2\rangle}.
	\label{e.RMSSprd}
\end{align}

%%%%%%%%%%%%%%%%%%%%%%%%%%%%%%%%%%%%%%%%%%%%%%%%%%%%%%%%%%%%%%%%%%%%%	

\subsection{Reliability}
\label{sec.reliability}

The RMS error is not always the appropriate measure to quantify the performance of an ensemble in probabilistic forecasting. For example, if the probability density function has disjoint support, the ensemble average may not have a physical meaning and can result in a poor forecast.  For probabilistic forecasts the reliability of an ensemble is more relevant. An ensemble is called {\it{perfectly reliable}} if the truth along with each ensemble member are independent draws from the same probability density function $\rho(X)$. In perfect ensembles the ratio between the RMS error and the RMS spread approaches $1$ as the ensemble size increases \citep{Wilks,LeutbecherPalmer08}. Under-dispersive ensembles, on the other hand, feature a ratio smaller than $1$, whereas over-dispersive ensembles feature ratios larger than $1$. Furthermore, in reliable ensembles the truth is statistically indistinguishable from any given ensemble member, and each ensemble member has equal probability to be closest to the truth. This property can be probed in Talagrand or Rank histograms \citep{Anderson96,HamillColucci97,Talagrand99}. For a given lead time, a Talagrand histogram is created by sorting the $N$ ensemble members in increasing order of their forecast value to form a set of $N+1$ bins. A histogram of probabilities of the truth falling into a bin $i$ at the given lead time is then produced by counting the frequency that the truth falls into the bin $i$. A reliable ensemble implies a flat histogram as the truth should have equal probability of falling into any given bin. Under-dispersive/over-dispersive ensembles, on the other hand, result in histograms which are convex/concave in shape \citep{Wilks}.

\section{Numerical results}
\label{sec.L96DA}

We now present numerical results demonstrating that SPBVs can be used as a reliable diverse ensemble with improved forecast skill in single-scale systems provided that BVs are not strongly localised. We shall present results for the strongly localised case with $K=128$ and for the weakly localised case with $K=40$ separately. We examine the ensemble diversity, forecast skill metrics such as the RMS error and RMS spread, as well as the reliability quantified by the error-spread relationship and the Talagrand diagram.\\

The setup for the numerical simulations is as follows. We employ an Ensemble Transform Kalman Filter to perform the data assimilation and construct the analysis  \citep{TippettEtAl03,WangEtAl04}. %Details are provided for completeness in the Appendix. 
The analysis is constructed from a forecast with a perfect model and noisy observations with variance $0.01$ (corresponding to observational noise with $2.75\%$ of the climatological standard deviation), following \cite{Bowler06,PazoEtAl13}. To focus on the performance of the bred vector ensemble rather than on the data assimilation, we use a large ensemble for the ETKF with $K+1$ members to produce the analysis, preventing filter divergence and avoiding the need for localisation and inflation. The ETKF ensemble is spun-up for $500$ time units before commencing the breeding cycles. The average analysis error for the $K=40$ and $K=128$ systems is $0.10$ and $0.18$, respectively, over $2,500$ forecasts. Ideally, the value of $\delta$ that results in a local minima of the RMS forecast error $\mathcal{E}$ matches the size of the analysis error.  In practice, however, the perturbation size often needs to be larger to achieve acceptable forecast skill \citep{TothKalnay97,MagnussonEtAl08,GigginsGottwald19}. 

We employ a breeding cycle length of $T=0.05$ time units. As is common practice in operational ensemble forecasting, pairs of positive/negative BVs are generated to ensure that the BV forecast ensemble represents the analysis mean at the initial forecast time. We consider BV ensembles consisting of $N$ BV perturbations of size $\delta$. For $K=40$ we use $N=10$ and for $K=128$ we use $N=20$ ensemble members, which implies $5$ and $10$ independent breeding cycles for $K=40$ and $K=128$, respectively. 
% Note that  for perturbation sizes $\delta$ for which ${\bar{\mathcal{D}}}_{ens}=1$ and BVs typically collapse onto the LLV, BVs generated from the analysis are the same as BVs generated from the truth. For larger values of $\delta$, when the alignment is only approximately valid on average, we find that BVs generated from the analysis and those generated from the truth are only statistically similar in the sense that they share the same ensemble dimension (not shown).\\
Each ensemble member is then evolved freely under the L96 dynamics for some lead time $\tau$. Forecasts are run for a total of $5$ time units and we report the results for lead times $\tau = 2.0$ and $\tau = 4.0$ time units for the forecast metrics presented in Section~\ref{sec.diagnostics}. A new forecast is created each $1.0$ time units. SPBVs have been generated using a noise strength of $\sigma = 1.25$. All metrics are averaged over $M=2,500$ forecasts.

\begin{figure}
	\centering
	\includegraphics[width=19pc]{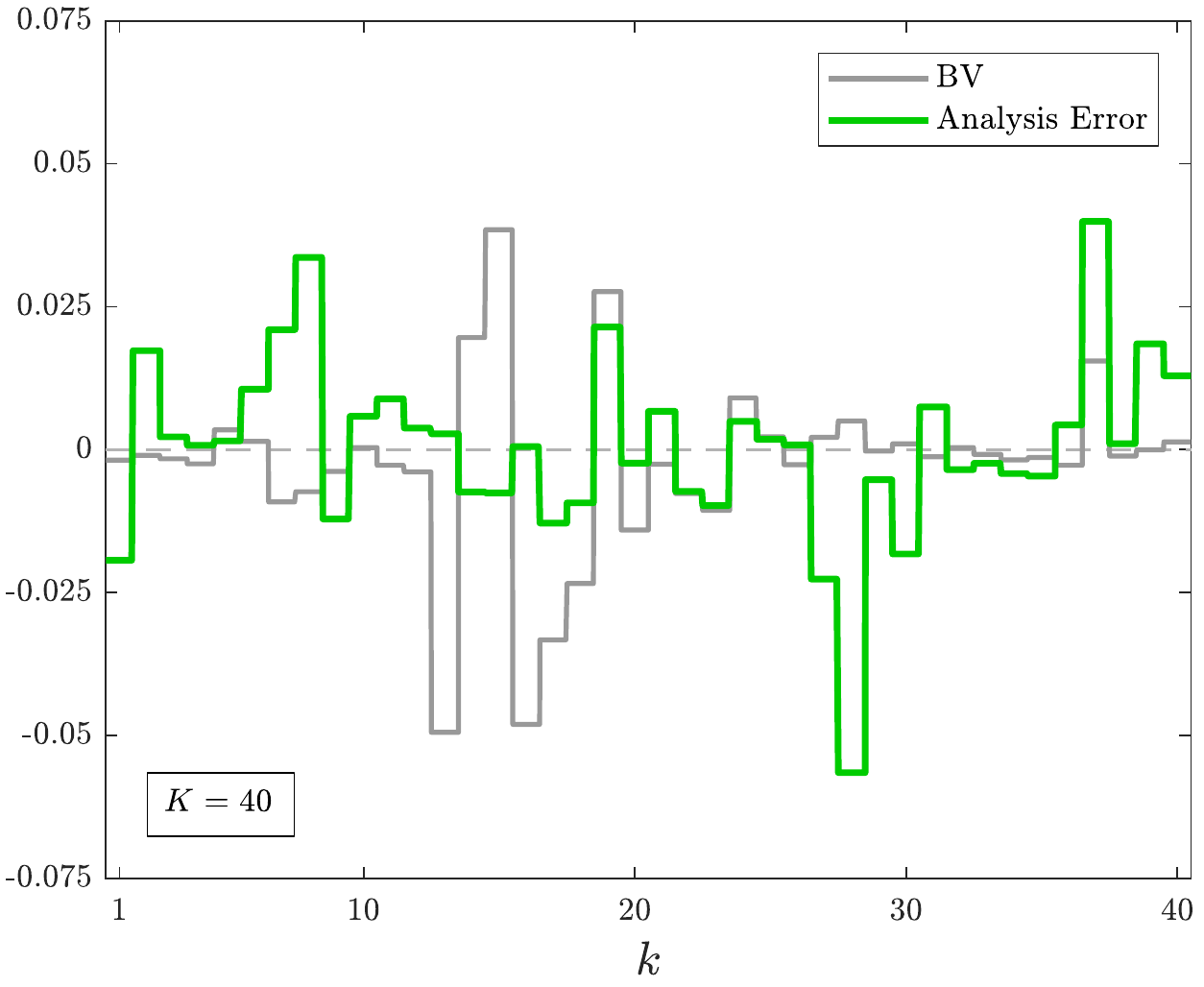}\\
	\includegraphics[width=19pc]{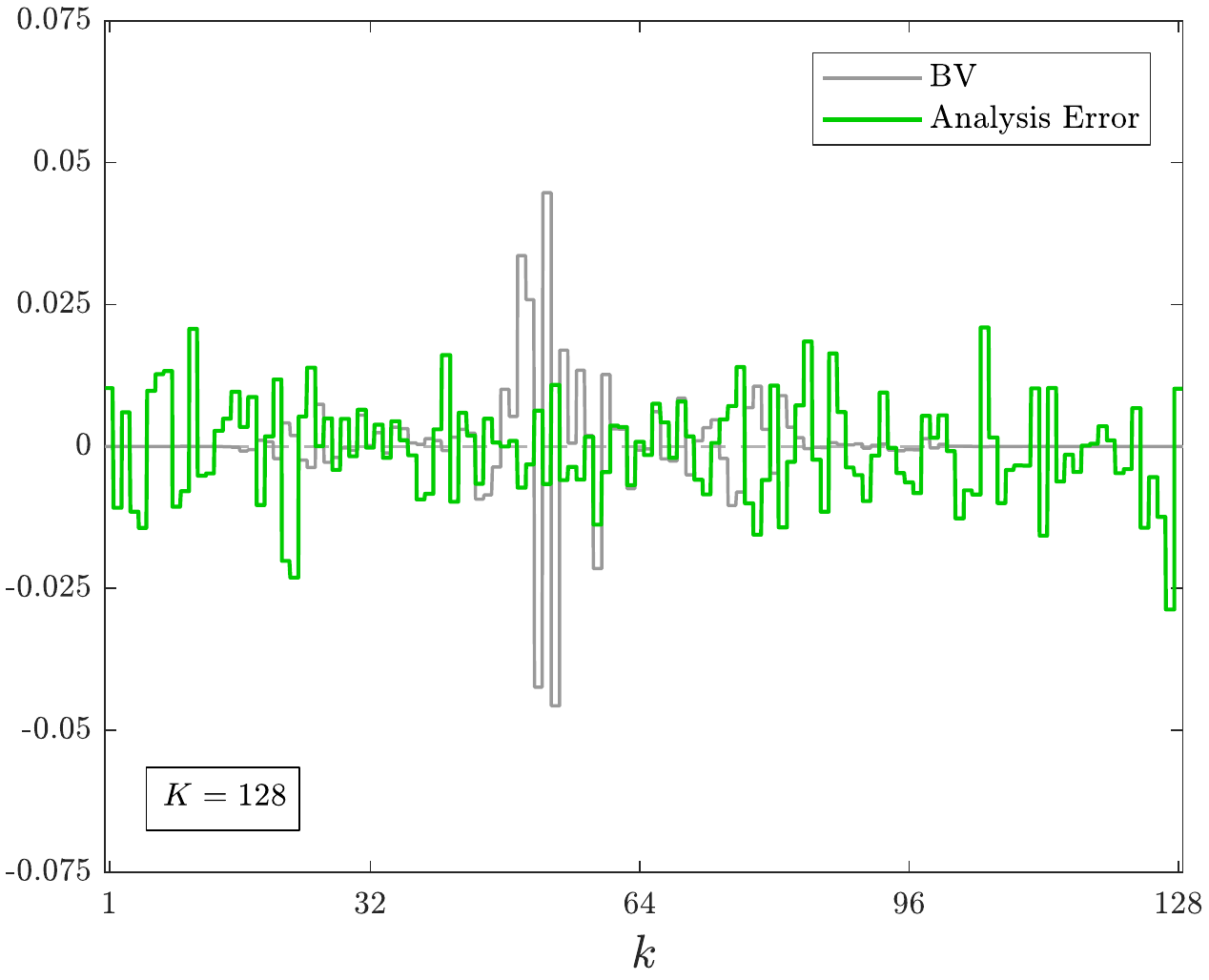}
	\caption{The bred vector from Figure~\ref{fig:L96BVSnapshot} and the analysis error at its time of generation. Top: $K=40$. Bottom: $K=128$.}
	\label{fig:L96AnalysisErrorSnapshot}
\end{figure}

Seeding BV forecast ensembles from the analysis fields poses problems, as a good forecast ensemble not only should evolve into likely future states but also has to account for the uncertainty of the analysis. This is particularly a problem in the case of strongly localised BVs (and SPBVs) for $K=128$, as the localisation inhibits sampling the uncertainty of the analysis which typically extends outside of the region of localisation and is distributed across the whole domain. Figure~\ref{fig:L96AnalysisErrorSnapshot} shows snapshots of BVs and the analysis error (scaled to have norm equal to $0.1$ to facilitate comparison) for the weakly localised case $K=40$ and the strongly localised case $K=128$. We remark that by construction, the spatial structure of SPBVs is similar to that of BVs. In the strongly localised case, it is clearly seen that there are large regions of significant uncertainty of the analysis which are not perturbed by the BV. In the weakly localised case $K=40$ on the other hand, BVs are more evenly distributed over the whole domain, and thus more likely to capture the errors in the analysis.
%The average localisation \eqref{e.localisationBV} of analysis errors is for $K=128$ $\bar{L}_{\textrm{a}} = 0.046$ (compared to $\bar{L}_{\textrm{BV}} = 0.15$ for BVs) quantifying this difference in localisation. For $K=40$ the difference in localisation is not significant with $\bar{L}_{\textrm{a}} = 0.11$ for analysis errors and ${\bar L}_{\textrm{BV}} = 0.16$ for BVs. 
%In the weakly localised case $K=40$ on the other hand, the average spatial structure of the analysis error, measured via the correlation matrix (\ref{e.BVCov}), is very well reproduced by all three bred vector ensembles, including RDBVs.

%\begin{figure}
%	\centering
%	\includegraphics[width=19pc]{./img/L96K40BVAnalysisCov.pdf}\\
%	\caption{Example row from the analysis error covariance $\P_a$ averaged over independent realisations. An example row of the covariance matrix \eqref{e.BVCov} for the BV displayed in Figure~\ref{fig:L96BVSnapshot} is depicted as a reference.}
%	\label{fig:L96AnalysisErrorCovariance}
%\end{figure}

The lack of activity in sites remote of the region of their spatial localisation is likely to severely inhibit the BV/SPBV ensemble to evolve into states which contain the truth. We shall find below, that the property of localisation is detrimental for the dynamically adapted SPBVs in the L96 system (\ref{e.L96}) without scale separation, whereas it was essential in the multi-scale case in \citet{GigginsGottwald19}. In particular we show that for $K=128$ strong localisation implies poor reliability of SPBV ensembles. RDBVs, however, despite not being dynamically adapted, exhibit improved reliability and forecast skill compared to classical BVs. In the weakly localised case $K=40$, SPBVs and RDBVs both constitute a reliable forecast ensemble with superior forecast skill compared to classical BV ensembles.

%%%%%%%%%%%%%%%%%%%%%%%%%%%%%%%%%%%%%%%%%%%%%%%%%%%%%%%%%%%%%%%%%%%%%	

\subsection{Ensemble Dimension}
\label{sec.DensResults}

\begin{figure}[h]
	\centering
	\includegraphics[width=19pc]{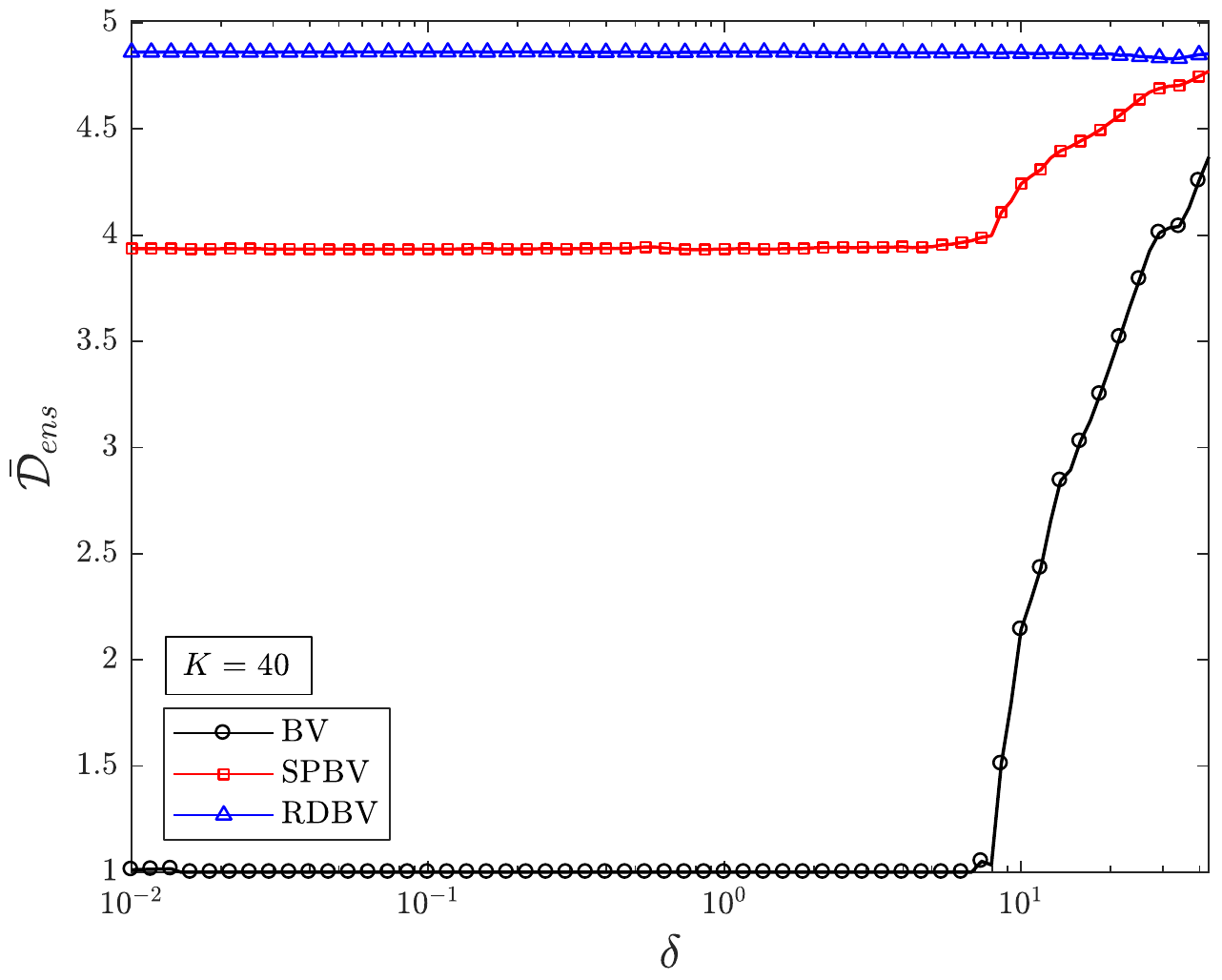}\\
	\vspace{1mm}
	\includegraphics[width=19pc]{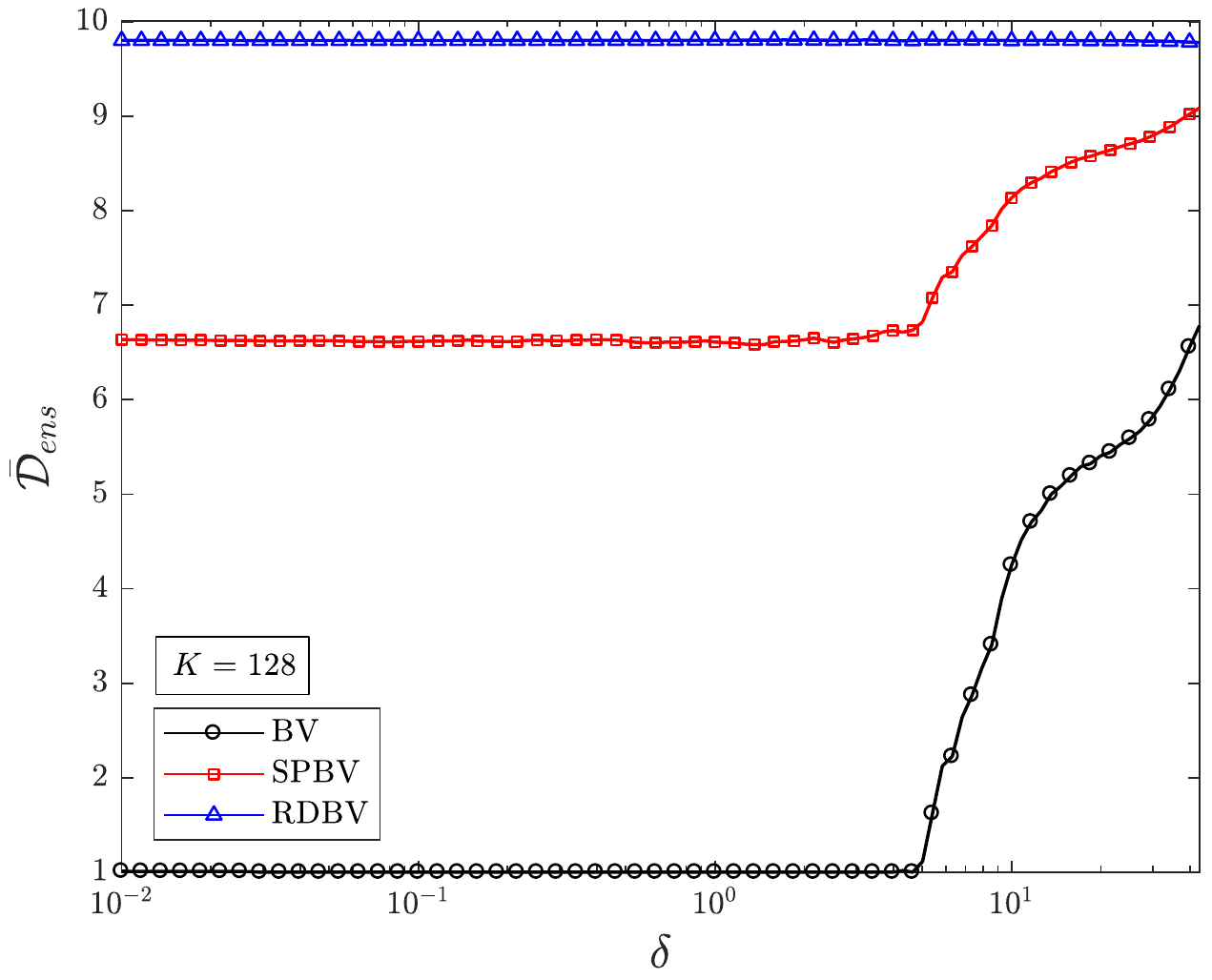}
	\caption{Average ensemble dimension ${\bar{\mathcal{D}}}_{ens}$ as a function of $\delta$ for each ensemble generation method for the L96 system \eqref{e.L96}. The SPBV ensemble was generated using $\sigma = 1.25$. 
	%${\bar{\mathcal{D}}}_{ens}$ was obtained as an average over $2500$ realisations. 
	Top: $K=40$. Bottom: $K=128$.}
	\label{fig:L96EnsDim}
\end{figure}

Figure~\ref{fig:L96EnsDim} shows the average ensemble dimension $\bar{\mathcal{D}}_{ens}$ \eqref{e.ensDimFormula} as a function of $\delta$ for classical BVs, SPBVs with $\sigma=1.25$, and RDBVs. For classical BVs the average ensemble dimension is ${\bar{\mathcal{D}}}_{ens} = 1$ for $\delta \lesssim 8$ and for $\delta \lesssim  5$, for $K=40$ and $K=128$ respectively, indicating the collapse of BV ensembles. For these perturbation sizes, a BV ensemble typically collapses onto the LLV but can also for $\delta>1$, when the dynamics of the perturbation begins to feel the nonlinearity of the dynamics, align in a different direction, spanned by the first few leading Lyapunov vectors (cf. Figures~\ref{fig:L96K40BLV}-\ref{fig:L96K128BLV}). For even larger perturbation sizes $\delta>8$ for $K=40$ and $\delta>5$ for $K=128$, the nonlinear dynamics becomes dominant and the ensemble dimension increases rapidly. Perturbation sizes corresponding to ${\bar{\mathcal{D}}}_{ens} >1$, however, are unrealistic in the sense that they are much larger than typical analysis errors for the L96 model as reported in \citet{Bowler06}, \citet{NgEtAl11} and \citet{PazoEtAl13}. This implies that classical BVs in our setting lack sufficient diversity. We remark that the qualitative behaviour of $\bar{\mathcal{D}}_{ens}$ does not change with the number of independent ensemble members $N$.\\

SPBVs and RDBVs exhibit a significant increase in the ensemble dimension. Both methods produce ensembles with a much larger ensemble dimension than the original BVs for all values of $\delta$. SPBVs maintain a consistent ensemble dimension of ${\bar{\mathcal{D}}}_{ens} = 3.9$ for $K=40$ and ${\bar{\mathcal{D}}}_{ens} = 6.6$ for $K=128$, before increasing in conjunction with the BVs when $\delta$ is large. RDBVs support the highest ensemble dimension as they are independent from each other. They do not attain the maximum ensemble dimension ${\bar{\mathcal{D}}}_{ens} = N$ since they are not strictly orthogonal. The averaged ensemble dimension of SPBVs is closer to the maximum ensemble dimension of ${{\mathcal{D}}}_{ens} =5$ for $K=40$  than to ${{\mathcal{D}}}_{ens} =10$ for $K=128$, reflecting the differing degree of localisation in the two cases; the multiplicative stochastic perturbation can generate a larger ensemble subspace the smaller the degree of localisation.\\ 

\begin{figure}[h]
	\centering
	\includegraphics[width=19pc]{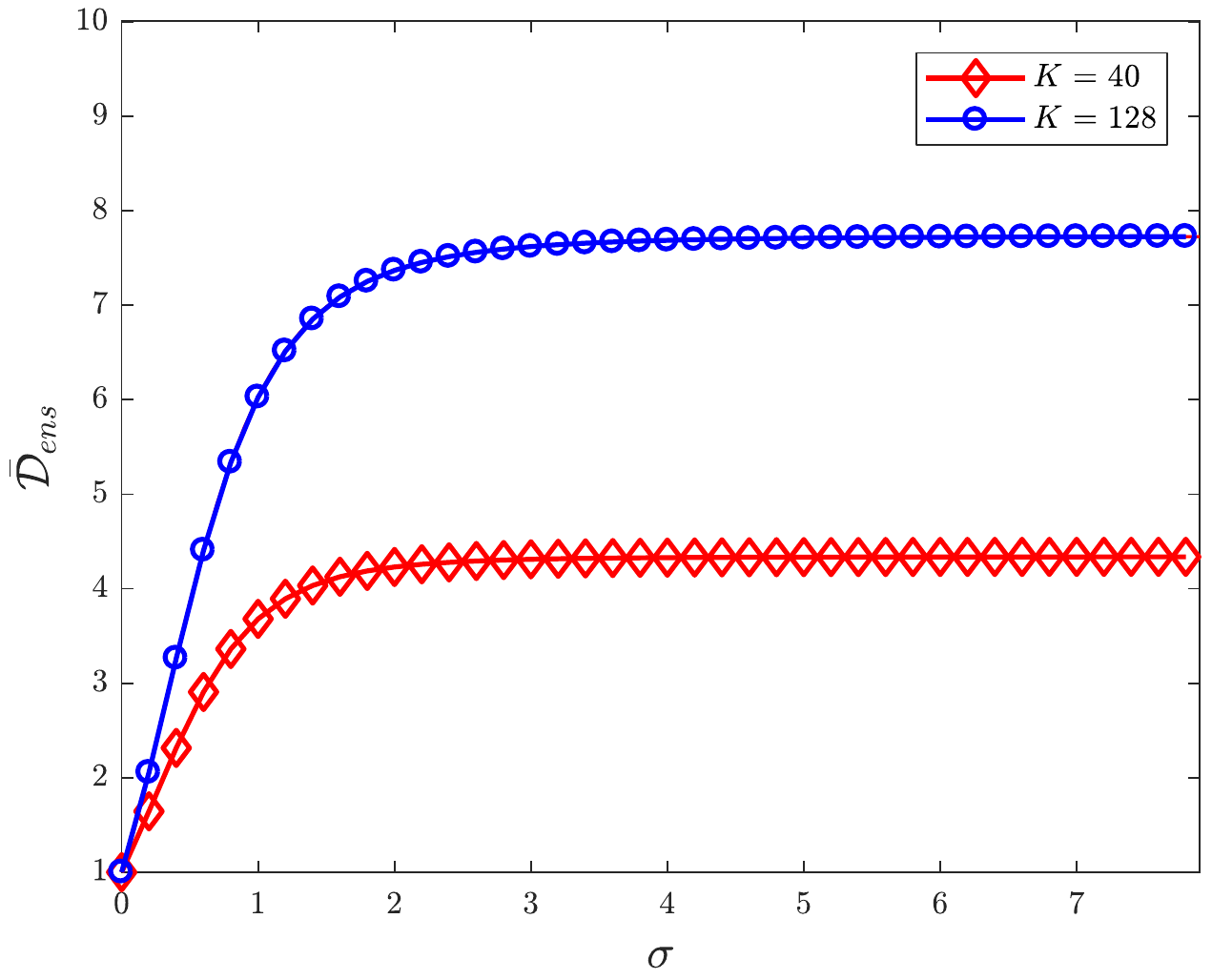}
	\caption{Average ensemble dimension ${\bar{\mathcal{D}}}_{ens}$ of an SPBV ensemble with perturbation size $\delta = 0.1$ as a function of $\sigma$ for the L96 system \eqref{e.L96} for $K=40$ (with $N=5$ ensemble members) and for $K=128$ (with $N=10$ ensemble members).}
	\label{fig:L96EnsDimSigma}
\end{figure}

The ensemble dimension of SPBVs increases for increasing values of the noise strength $\sigma$ as shown in Figure~\ref{fig:L96EnsDimSigma} for SPBVs with $\delta=0.1$. The ensemble dimension approaches a limiting value of ${\bar{\mathcal{D}}}_{ens} = 4.3$ for $K=40$ and of ${\bar{\mathcal{D}}}_{ens} = 7.7$ for $K=128$. The limiting ensemble dimension is smaller than $N$ for both $K=40$ and $K=128$. The difference is larger for the strongly localised case $K=128$ for the same reason as discussed above. The observed increase of the ensemble dimension with increasing noise strength may suggest that one should use sufficiently large noise strengths $\sigma$ and moreover that the performance is insensitive to changes in $\sigma$ past some threshold value. We will see that this is correct for the forecast skill and the reliability. However, as we have seen in Section~\ref{sec.CLV} dynamic adaptivity is lost for too large values of the noise strength $\sigma$.
% but that increasing the noise strength leads to unreliable over-dispersive ensembles. 

\subsection{Ensemble Forecast Skill}
\label{sec.L96SkillDA}

\begin{figure}
	\centering
	\includegraphics[width=19pc]{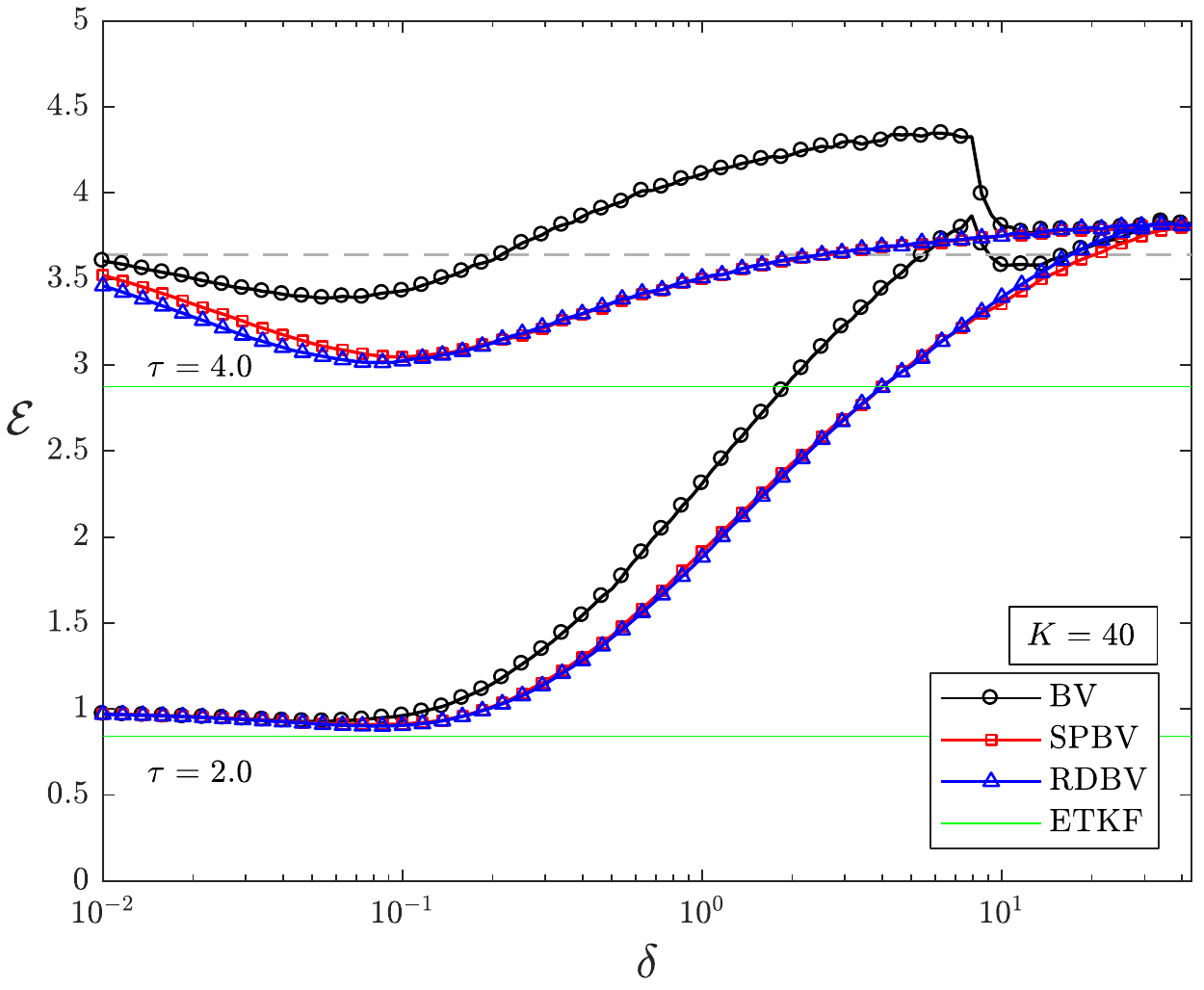}\\
	\vspace{1mm}
	\includegraphics[width=19pc]{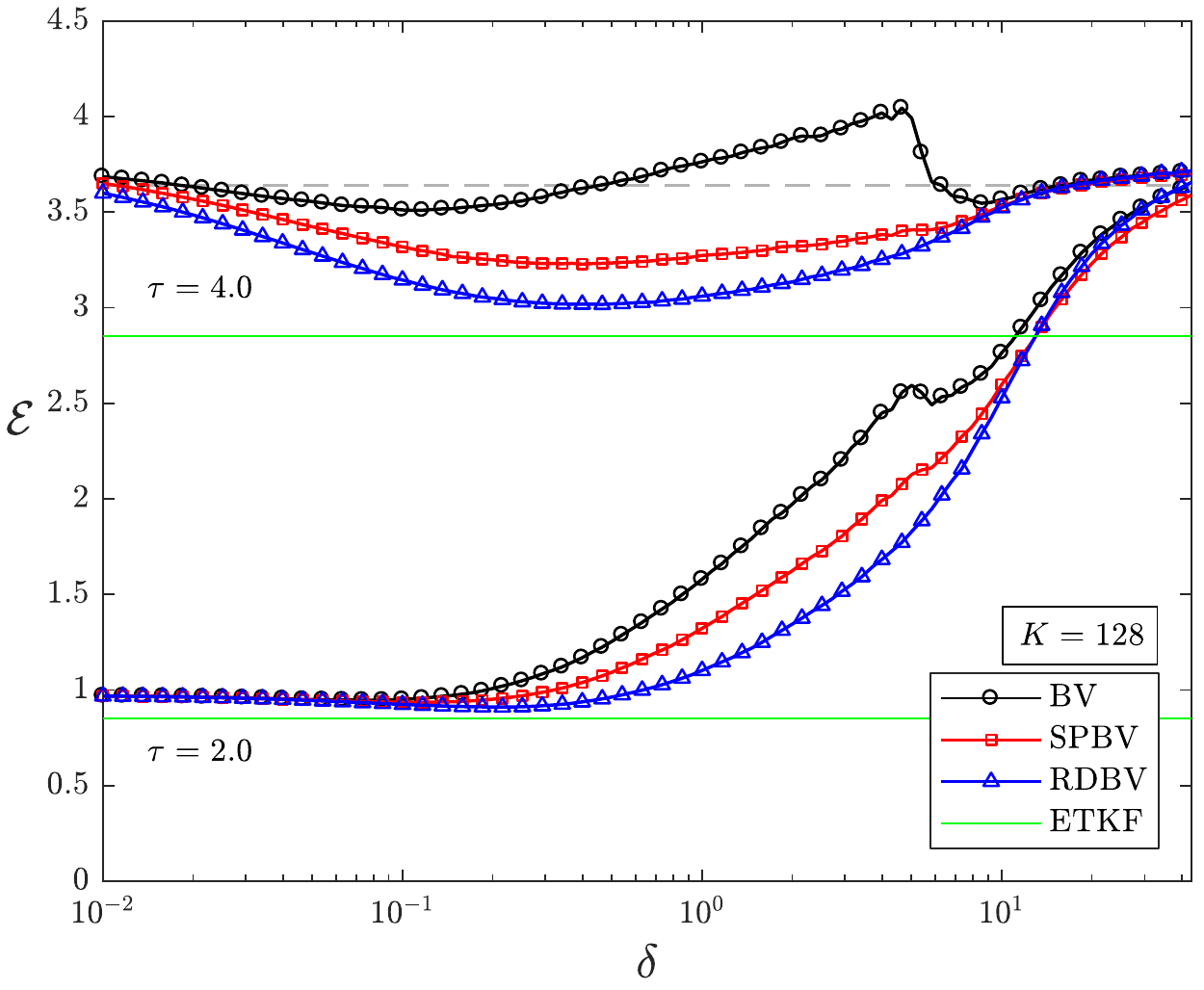}
	\caption{RMS error $\mathcal{E}$ as a function of $\delta$ for each ensemble generation method for fixed lead times $\tau = 2.0$ and $\tau = 4.0$. The dashed reference lines denote the climatic error $\mathcal{E}_{\rm{clim}}$ and the RMS error of an ETKF ensemble with $K+1$ ensemble members at the respective lead times. Top: $K=40$. Bottom: $K=128$.}
	\label{fig:L96RMSErrDA}
\end{figure}

Figure~\ref{fig:L96RMSErrDA} shows the RMS error $\mathcal{E}$ for BVs, SPBVs and RDBVs as a function of the perturbation size $\delta$ for lead times $\tau = 2.0$ and $\tau = 4.0$, both for the strongly localised case $K=128$ and the weakly localised case $K=40$. We also show as reference the climatic error $\mathcal{E}_{\rm{clim}}=\sigma_{\rm{clim}}\approx 3.64$ as well as the RMS error of an ETKF ensemble with a larger ensemble size of $N=K+1=41$ and $N=K+1=129$, respectively, to provide an upper bound for the forecast skill.

%We remark that the ETKF ensemble has ensemble size of $N=K+1$ significantly larger than the ensemble sizes of the respective BV ensembles, and serves as a gold standard here rather than to compare BV ensembles to ETKF ensembles.

%It is seen that the RMS error exhibits a local minimum for a designated perturbation size $\delta_{\rm{min}}$. For $K=128$ we find $\deltamin=???$ for BVs, $\deltamin=???/???$ for SPBVs and $\deltamin=???/???$ for RDBVs for lead times $\tau=2.0/4.0$. For $K=128$ we find $\deltamin=???/???$ for BVs, $\deltamin=???/???$ for SPBVs and $\deltamin=???/???$ for RDBVs for lead times $\tau=2.0/4.0$. Hence all bred vector ensembles have their minimal error at perturbation sizes consistent with the average analysis error of $0.10$ and $0.18$ for the $K=40$ and $K=128$.\\
%\gaginline{Is this correct? Please provide the numbers.}
Classical BVs exhibit the largest RMS error for both cases and both lead times and for all values of $\delta$. The RMS error exhibits a local minimum for a designated perturbation size $\delta_{\rm{min}}$. For $K=40$ we find $\deltamin=0.05$ for BVs for both lead times $\tau=2.0$ and $\tau=4.0$. For $K=128$ we find for BVs $\deltamin=0.08$ and $\deltamin=0.12$ for lead times $\tau=2.0$ and $\tau=4.0$, respectively. Hence BV ensembles exhibit their minimal RMS error at perturbation sizes which are not consistent with the average analysis error of $0.10$ and $0.18$ for $K=40$ and $K=128$, respectively.  For perturbation sizes around $\deltamin$, BVs are collapsed to a single member with ensemble dimension ${\bar{\mathcal{D}}}_{ens} = 1$ at both lead times. The RMS error assumes unacceptable high values for perturbation sizes which allow for a non-collapsed BV ensemble with ${\bar{\mathcal{D}}}_{ens} > 1$ for $\delta\gtrsim8$ ($\delta\gtrsim5$ for $K=128$).\\

%Hence, we see that the minimum BV RMS error for both lead times $\tau = 2.0$ and $\tau = 4.0$ occur at values of $\delta$ close to the expected analysis error, but the value of the RMS error is larger than that of the other ensemble methods due to having ${\bar{\mathcal{D}}}_{ens} = 1$ at these perturbation sizes. Similarly to the unrealistic truth experiment presented in Section~\ref{sec.L96tr}, for large values of $\delta\gtrsim8$ ($\delta\gtrsim5$ for $K=128$), the RMS error is still large despite having ${\bar{\mathcal{D}}}_{ens} > 1$ and does not perform better than BV ensembles with $\delta$ matching the analysis error for all lead times. \\

% The qualitative behaviour of the error of BVs is similar to the case of the unrealistic truth experiment presented in Section~\ref{sec.L96tr}. For $\delta>8$ ($\delta>3$) the RMS error approaches the climatological error $\mathcal{E}_{\rm{clim}}=\sigma_{\rm{clim}}\approx 3.64$ for $K=40$ ($K=128$), and for perturbation sizes when the ensemble dimension is ${\bar{\mathcal{D}}}_{ens} = 1$, indicating ensemble collapse, the collapsed ensemble approaches again $\sqrt{2}\mathcal{E}_{\rm{clim}}\approx 5.2$. For $\tau=4$ BVs do not achieve any skill and perform worse than the climatological mean error $\mathcal{E}_{\rm{clim}}$ for all values of the perturbation size $\delta$. 

% The poor performance in forecast skill for small values of $\delta$ stems from the lack of diversity with ${\bar{\mathcal{D}}}_{ens}=1$ and the collapse (on average) onto the LLV (cf. Figure~\ref{fig:L96EnsDim}). 

SPBV and RDBV ensembles exhibit a significant increase in forecast skill and consistently have smaller RMS error for all values of the perturbation size $\delta$ and lead times. For the smaller lead time $\tau=2$ we observe that all ensembles incur the same RMS error for sufficiently small values of the perturbation size $\delta$. This is because for perturbation sizes $\delta$ which are significantly smaller than the analysis error, all ensembles are highly under-dispersive and the forecast error is dominated by the analysis error. SPBV and RDBV ensembles perform almost identically in the weakly localised case $K=40$, due to their similar spatial structures, and both ensembles perturb significantly across the whole domain capturing the regions of non-trivial analysis error. Both the SPBV and RDBV ensembles feature an RMS error minimum at approximately $\delta = 0.1$ for all lead times, which matches the size of the average analysis error, almost attaining the forecast skill of an ETKF ensemble with a much larger ensemble size of $N=41$. We have checked that the minimal RMS error approaches the reference value provided by the ETKF for increasing ensemble sizes of SPBVs and RDBVs. This indicates that the ensembles are well-adapted to capturing the analysis error uncertainties in addition to capturing the dynamic error growth. In the strongly localised case $K=128$, RDBVs consistently exhibit smaller forecast RMS errors compared to SPBVs. In the strongly localised case, the optimal perturbation size associated with the smallest RMS error depends on the lead time $\tau$ for SPBVs and RDBVs. For lead times $\tau=2.0$ we find $\deltamin=0.18$ for SPBVs and $\deltamin=0.22$ for RDBVs, consistent with the average analysis error of $0.18$. The RMS error of RDBVs approaches the reference value provided by the ETKF ensemble for $\tau=2.0$. For $\tau=4.0$ we find $\deltamin=0.39$ for SPBVs and $\deltamin=0.46$ for RDBVs, which are both inconsistent with the average analysis error. Hence, in the strongly localised case the optimal perturbation size $\deltamin$ of SPBVs and RDBVs does not match the average analysis for all lead times. This suggests that we may not be efficiently capturing the uncertainties of the analysis. The difference between the ETKF reference ensemble and the BV ensembles for $K=128$ is larger than that for $K=40$ due to the aforementioned strongly localised nature of the BV ensemble perturbations.\\

%The difference is larger for $K=128$ where the relative difference in ensemble size between bred vector ensembles with $N=10$ and ETKF ensembles with $N=129$ is larger than for $N=40$ with $N=10$ for bred vector ensembles and $N=41$ for ETKF ensembles.
%The RMS error of RDBV at $\delta_{\rm{min}}$, however, is only slightly larger than the reference value provided by the ETKF ensemble.\\

We now discuss how the RMS error for SPBVs changes as the noise strength $\sigma$ is varied. We recall that when $\sigma \to 0$ SPBVs essentially reproduce the original BVs they were generated from, while once $\sigma$ is sufficiently large, the ensemble dimension saturates at some fixed value due to the rescaling back to size $\delta$ (cf. Figure~\ref{fig:L96EnsDimSigma}). The RMS error of SPBVs deviates rapidly from the value attained by BVs for increasing values of $\sigma$, and then asymptotes to a constant value for large $\sigma$ (not shown). For the weakly localised case $K=40$ the asymptotic value of the RMS error of SPBV is close to the one of RDBVs - the spatial structure of both ensembles is not related to the current state and their associated initial conditions evolve into random draws from the attractor, so both ensembles have the same statistical properties. For the strongly localised case $K=128$, on the other hand, the asymptotic RMS error of SPBVs is larger than the one of RDBVs. In the localised case, SPBV ensembles have markedly different statistical properties to RDBVs as they sample locally with all ensemble members exhibiting non-vanishing entries in the same spatial region. We found that increasing $\sigma$ past $\sigma=1.25$ does not increase the forecast skill in terms of RMS error for the lead times considered here.\\

\begin{figure}
	\centering
	\includegraphics[width=19pc]{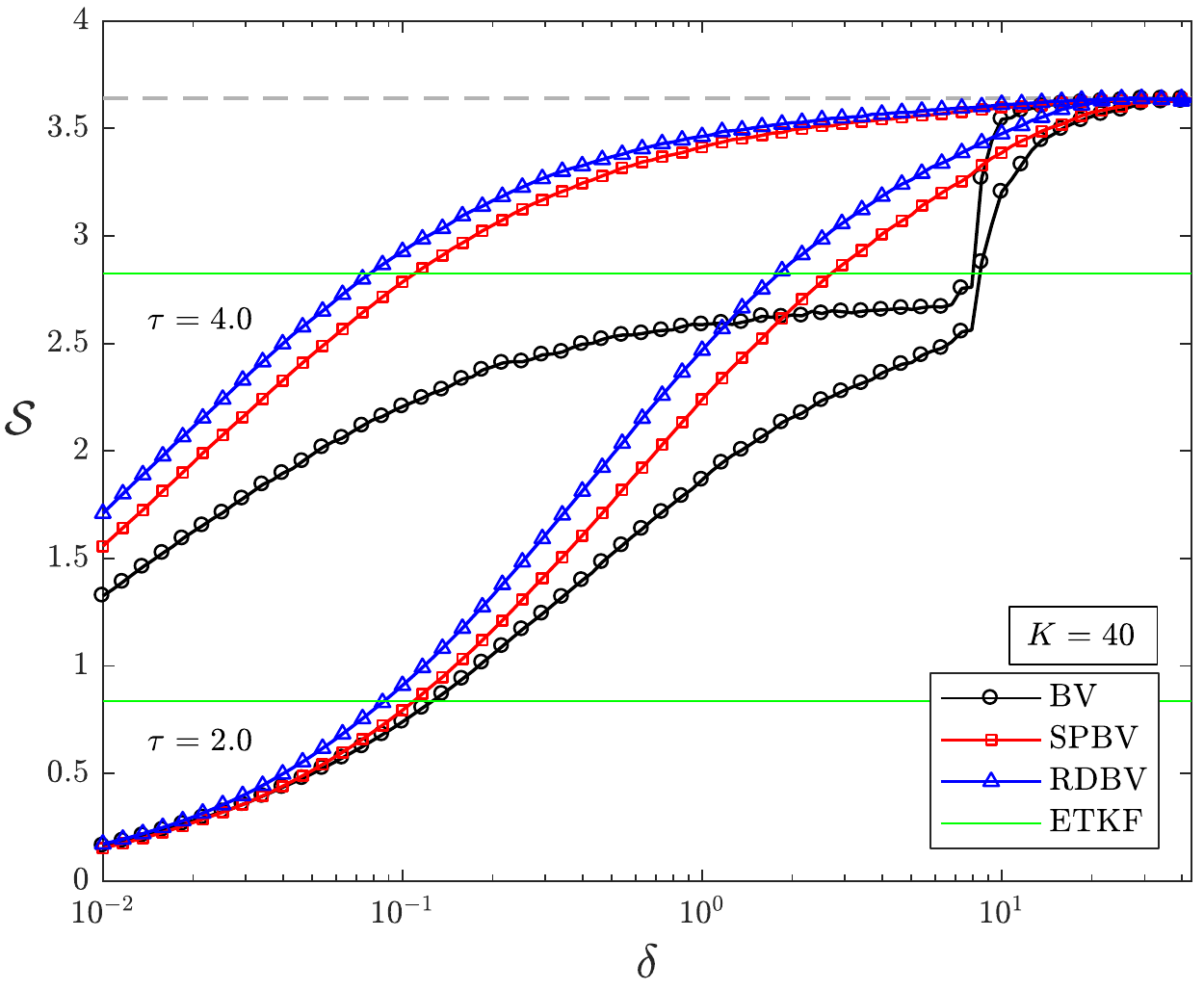}\\
	\includegraphics[width=19pc]{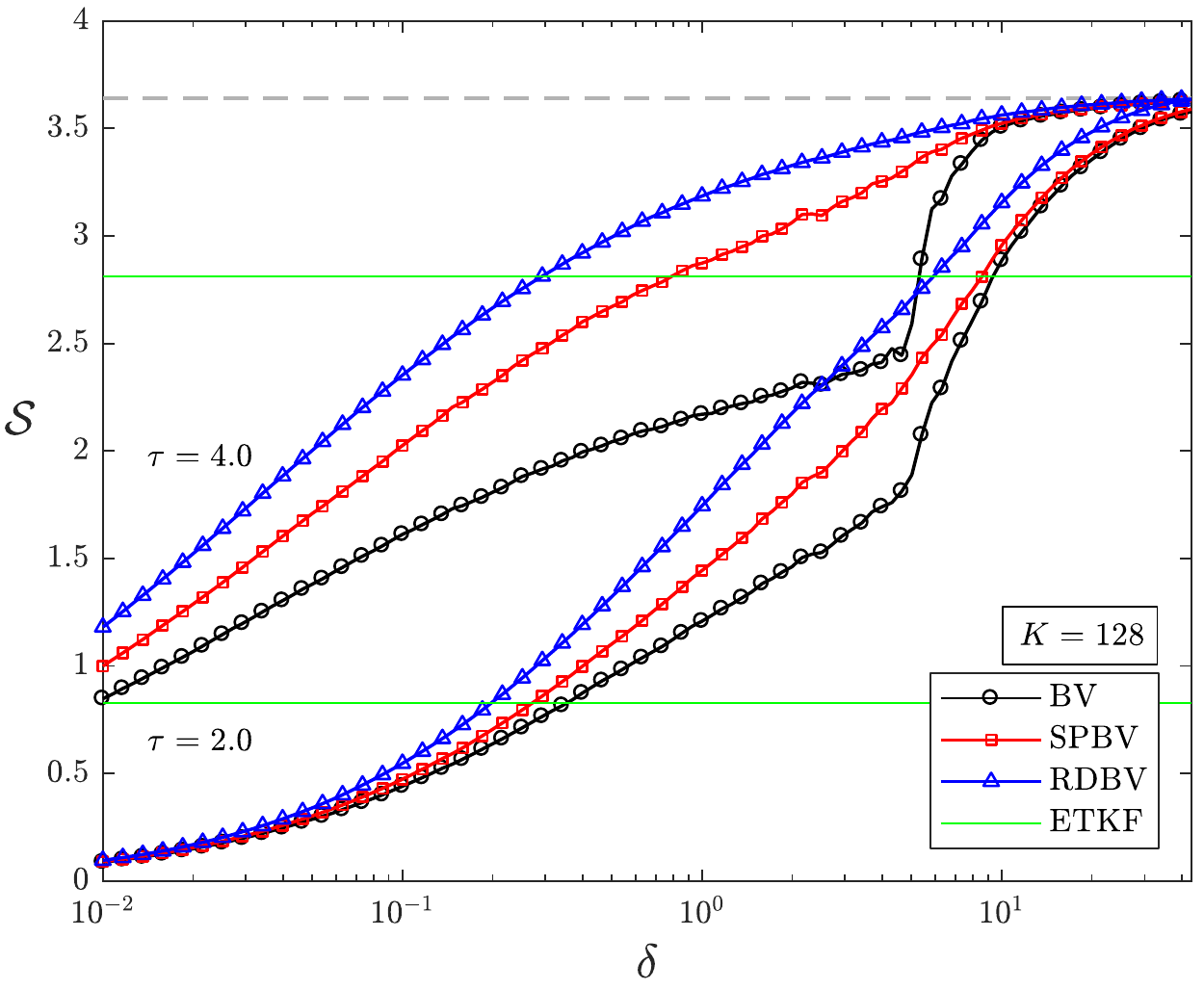}
	\caption{RMS spread $\mathcal{S}$ for forecast ensembles as a function of $\delta$ for each ensemble generation method for fixed lead times $\tau = 2.0$ and $\tau = 4.0$. The dashed reference lines denote the climatic error $\mathcal{E}_{\rm{clim}}$ and the RMS error of an ETKF ensemble with $K+1$ ensemble members at the respective lead times. Top: $K=40$. Bottom: $K=128$.}
	\label{fig:L96RMSSprdDA}
\end{figure}

The ensemble RMS spread $\mathcal{S}$ \eqref{e.RMSSprd} for BVs, SPBVs and RDBVs as a function of the perturbation size $\delta$ is shown in Figure~\ref{fig:L96RMSSprdDA}. For reference we also depict the corresponding RMS spread of an ETKF ensemble. The results are consistent with those of the ensemble dimension and of the RMS error $\mathcal{E}$ shown above. It is clearly seen that classical BVs are deficient in RMS spread. Classical BV ensembles exhibit a non-vanishing spread for small values of $\delta$ despite their ensemble dimension being only ${\bar{\mathcal{D}}}_{ens} = 1$. This is entirely due to the chosen set-up of using pairs of positive and negative BVs, and is not indicative of any non-trivial diversity of the ensemble. Once ${\bar{\mathcal{D}}}_{ens} > 1$ (cf. Figure~\ref{fig:L96EnsDim}) the RMS spread of BVs increases significantly. SPBVs and RDBVs exhibit significantly larger RMS spread compared to BVs. As for the RMS error, the differences between the RMS spread of RDBVs and SPBVs is more pronounced in the strongly localised case $K=128$, reflecting the reduced ensemble space of SPBVs which are generated from a collapsed strongly localised BV by multiplicative perturbations, preserving the localisation. In the weakly localised case $K=40$ the smaller RMS spread of SPBVs compared to RDBVs implies that SPBVs achieve the same forecast skill with less ensemble spread. In the strongly localised case $K=128$ the increased ensemble spread of RDBVs positively impacts on their forecast skill and their RMS error.

%%%%%%%%%%%%%%%%%%%%%%%%%%%%%%%%%%%%%%%%%%%%%%%%%%%%%%%%%%%%%%%%%%%%%

\subsection{Reliability}
\label{sec.L96ReliabilityDA}

\begin{figure}
	\centering
	\includegraphics[width=19pc]{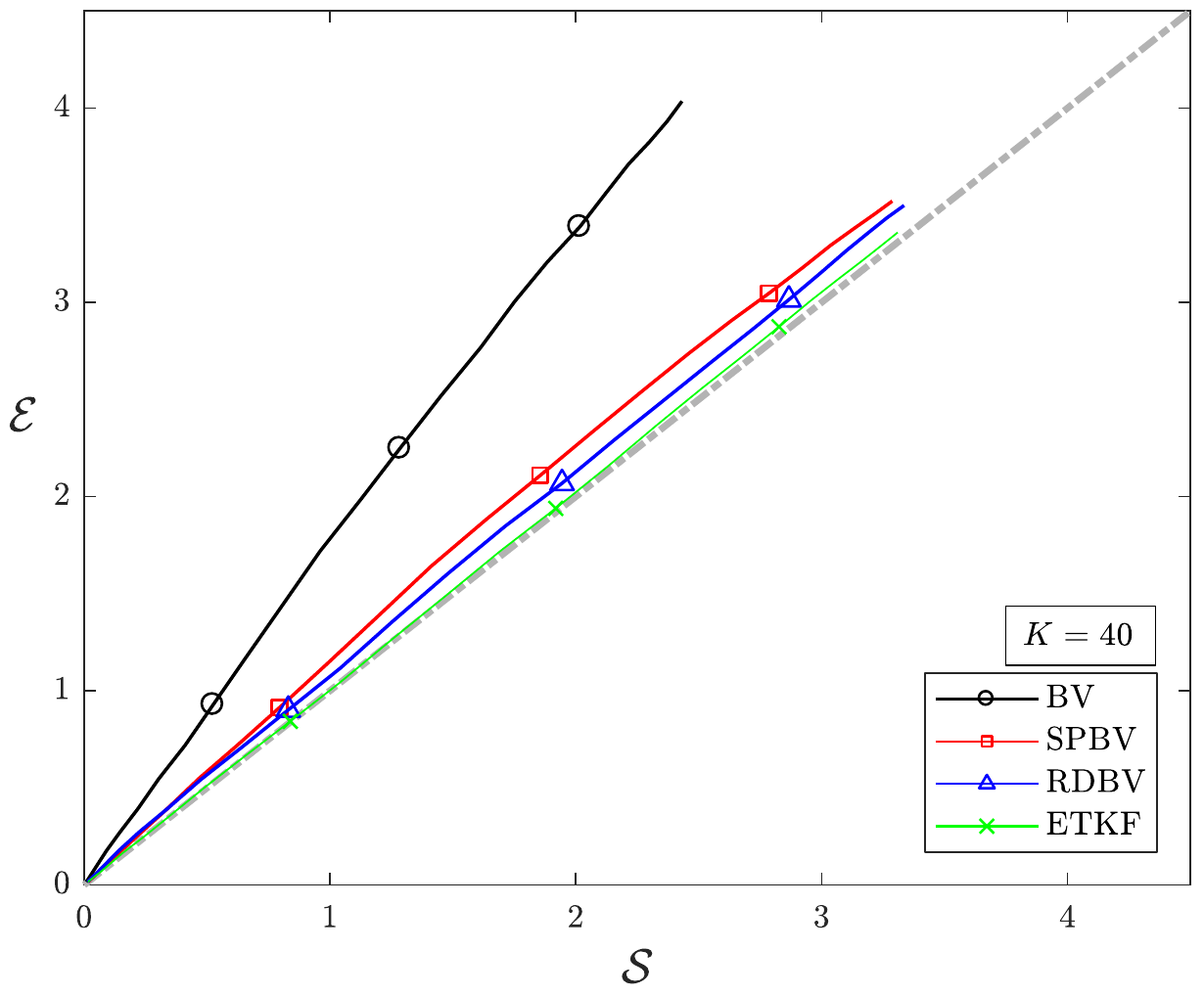}\\
	\vspace{1mm}
	\includegraphics[width=19pc]{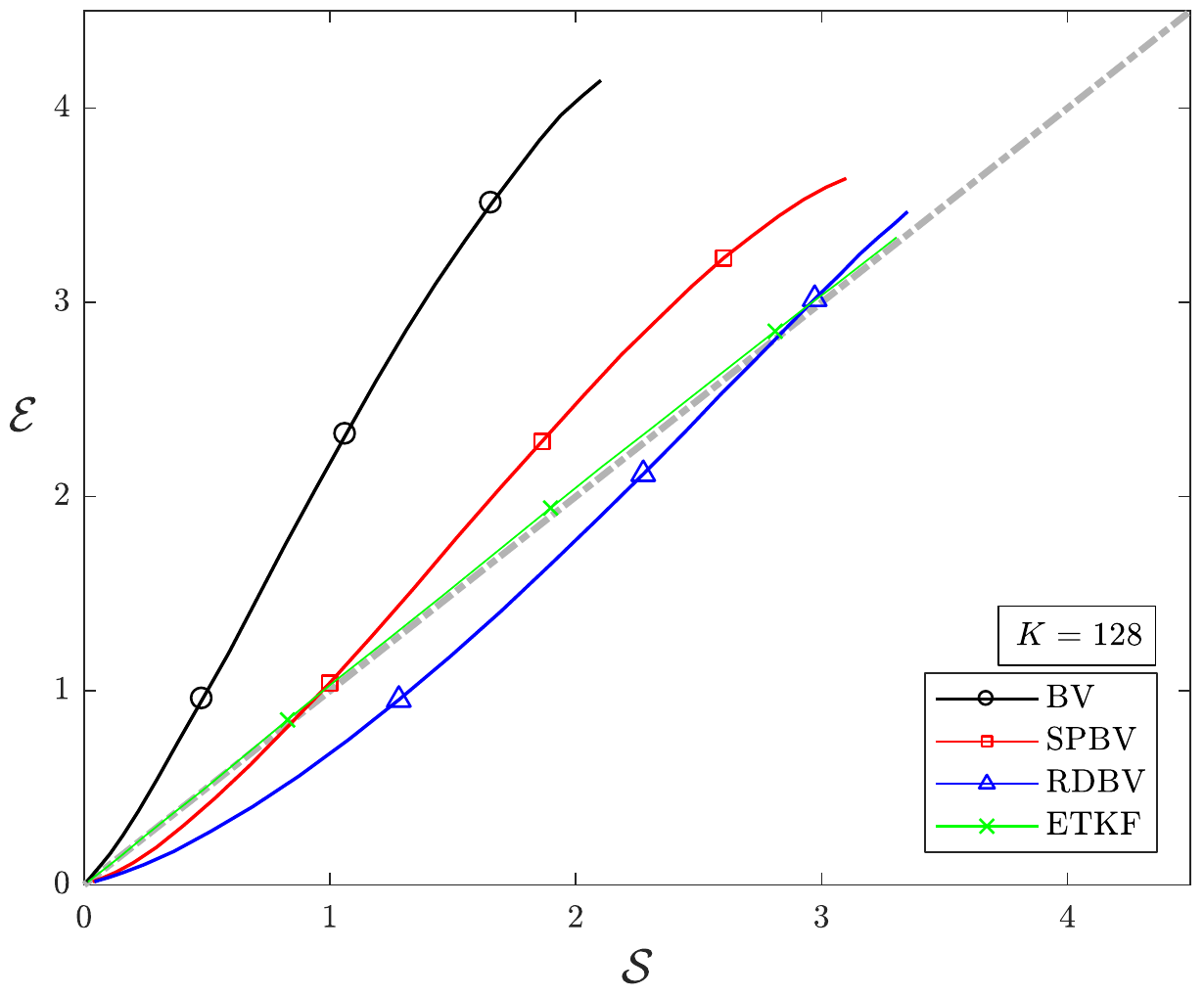}
	\caption{RMS error vs RMS spread, parameterised by increasing lead times from $\tau=0$ to $\tau=5.0$. Each bred vector ensemble was generated using a perturbation sizes $\delta_{\rm{min}}$ corresponding to minimal RMS error for each ensemble type at lead time $\tau=4.0$ (cf. Figure~\ref{fig:L96RMSErrDA}). The markers indicate the specific lead times $\tau = 2.0$, $\tau = 3.0$ and $\tau = 4.0$. The grey dot-dashed line indicates a one-to-one ratio of RMS error and RMS spread, corresponding to a reliable ensemble. Top: $K=40$. Bottom: $K=128$.}
	\label{fig:L96ErrSprdDA}
\end{figure}

We now use the error-spread ratio and the Talagrand histogram to evaluate if the additional spread acquired by the stochastic modifications of BVs is beneficial in the sense that it leads to a reliable ensemble or whether it causes the ensemble to be simply over-dispersive. The error-spread ratio, parameterised by lead time $\tau$, is shown in Figure~\ref{fig:L96ErrSprdDA}. The markers indicate the lead times of $\tau = 2.0$, $\tau = 3.0$ and $\tau = 4.0$. The reference ETKF ensembles show a reliable one-to-one ratio. For the bred vectors and their stochastic modifications, each curve was obtained using a different value of the perturbation size $\delta$, which corresponds to the optimal perturbation size producing the smallest RMS error at lead time $\tau=4.0$ (cf. Figure~\ref{fig:L96RMSErrDA}). For the weakly localised case $K=40$ both SPBVs and RDBVs are close to the ideal error-spread ratio of $1$, suggesting a reliable ensemble. Since the ensemble size $N=10$ is relatively small, the error-spread curves lie just above the one-to-one ratio due to finite-size sampling error effects. On the other hand, for the strongly localised case $K=128$ SPBV ensembles are over-dispersive for small lead times $\tau\le 2.0$, becoming under-dispersive for lead times $\tau>2.0$. RDBV ensembles are seen to be over-dispersive for all lead times $\tau \le 4.0$.\\

\begin{figure}
	\centering
	\includegraphics[width=19pc]{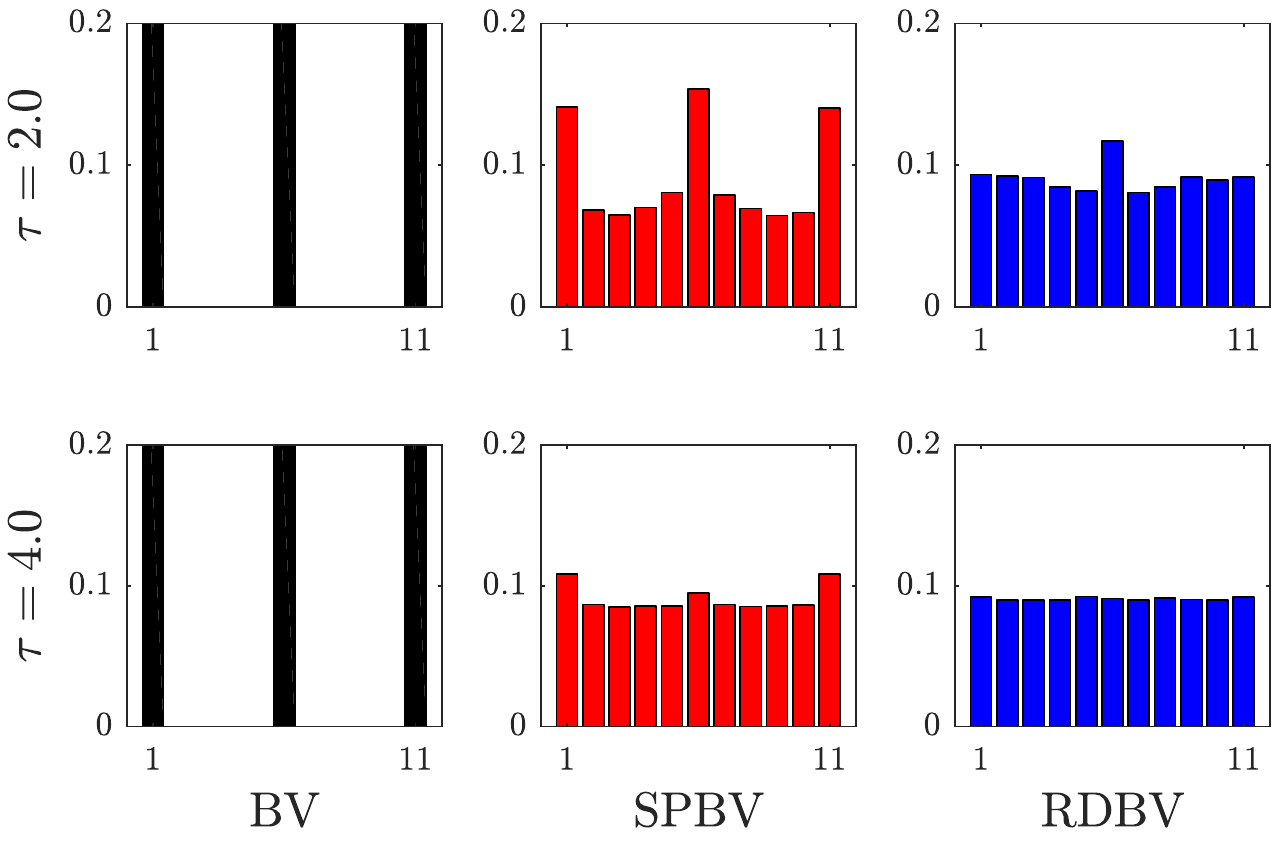}\\
	\includegraphics[width=19pc]{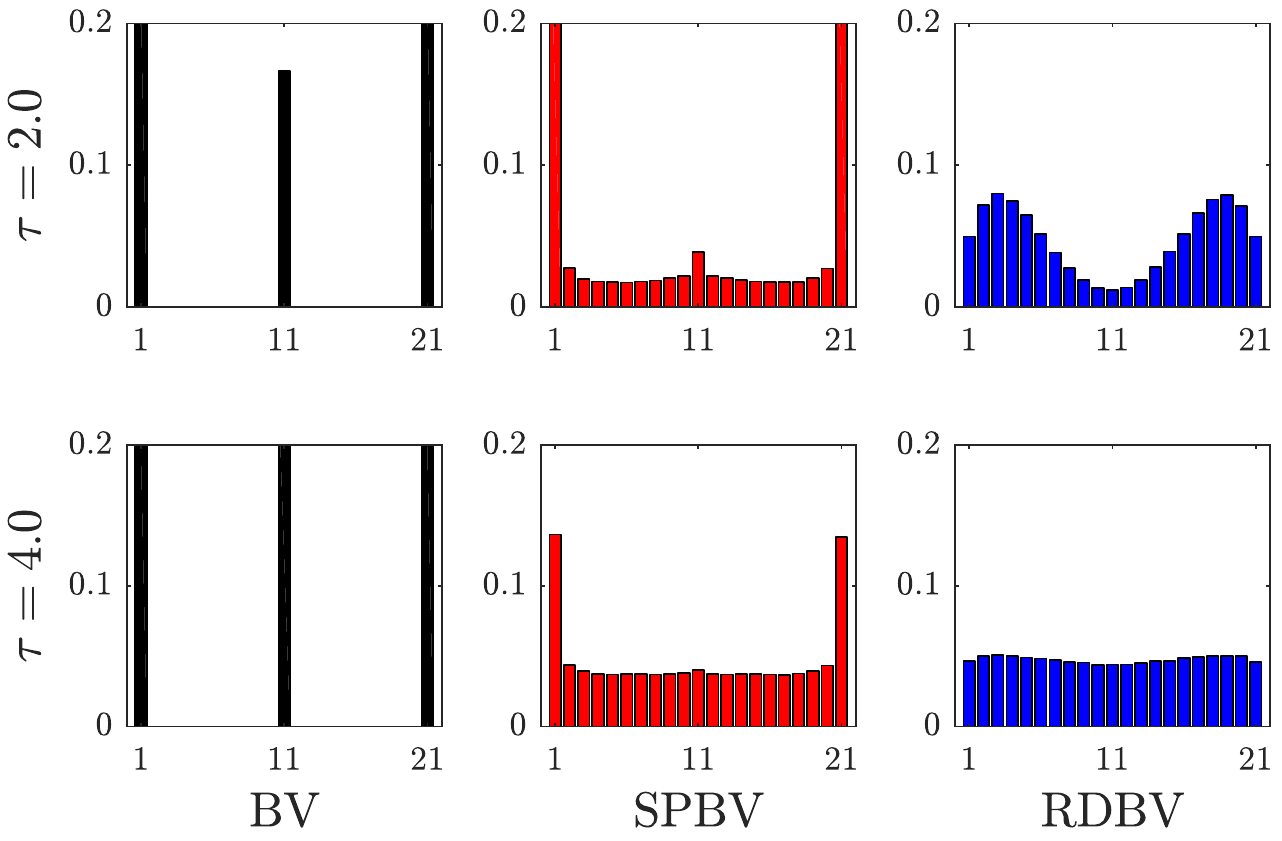}
	\caption{Talagrand diagrams for forecast ensembles for lead times $\tau=2.0$ and $\tau=4.0$. Each ensemble was generated using a perturbation sizes $\delta_{\rm{min}}$ corresponding to minimal RMS error for each ensemble type and lead time (cf. Figure~\ref{fig:L96RMSErrDA}). Top: $K=40$. Bottom: $K=128$.}
	\label{fig:L96TalHistDA}
\end{figure}

Talagrand histograms are shown for each of the three forecast ensembles, averaged over all sites, in Figure~\ref{fig:L96TalHistDA}, for lead times $\tau=2.0$ and $\tau=4.0$. Each histogram was again obtained using the perturbation size $\delta$ corresponding to the respective minimal RMS error (cf. Figure~\ref{fig:L96RMSErrDA}). Consistent with the results on the error-spread ratio above, the Talagrand diagrams show that SPBV and RDBV ensembles are reliable with a flat histogram in the weakly localised case $K=40$. The reliability of the stochastically modified BV ensembles is linked to the fact that they generate non-trivial variance in regions of non-vanishing analysis error. Consistent with the observed perfect one-to-one error-spread ratio, ETKF ensembles exhibit a flat Talagrand diagram (not shown).  

On the other hand, in the strongly localised case $K=128$, when there is a strong discrepancy between the spatial structure of the analysis error and all of the SPBV ensemble members, SPBV ensembles do not lead to a flat Talagrand histogram, indicating an unreliable under-dispersive ensemble. We observe a high probability for the true state to lie outside the ensemble for both lead times. Remarkably, the interior bins of the histogram are relatively evenly populated and we do not observe the ``U" shape typically associated with under-dispersive ensembles. The unusual shape of the Talagrand histogram in the strongly localised case with a flat region embedded between two peaks can be understood as follows. Consider an arbitrary component $k$ of an SPBV away from the localised region of the parent BV, which is not significantly perturbed. If ${\bar{\mathcal{D}}}_{ens} = 1$, then none of the members of the SPBV ensemble will be able to perturb this site. The initial conditions associated with these SPBVs at site $k$ are therefore approximately equal to the analysis mean at that site. However, typically the true state is much further away in phase space from the analysis mean. After evolving the SPBV ensemble forward in time, for reasonable lead times the ensemble has likely not developed sufficient spread to enclose the truth within its support. Hence, in the corresponding Talagrand diagram the truth falls into one of the exterior bins. This explains the peaks at the edge of the Talagrand histogram of SPBVs observed in Figure~\ref{fig:L96TalHistDA}. On the other hand, the non-trivial components of an SPBV corresponding to the localised region have comparable magnitude to that of the analysis errors. This ensures that there are several components of the L96 model for which the true state is contained within the ensemble, contributing to the evenly distributed tally marks in the middle of the Talagrand histogram. 

RDBV ensembles display an unusual shape of the rank histogram for $\tau = 2.0$ with two distinct modes in the strongly localised case $K=128$.  This is again linked to the mismatch between the spatial structure of localised BVs and the analysis error.  Individual RDBV ensemble members do not efficiently sample the analysis error since each individual RDBV is localised. On the other hand it is likely that each site will be significantly perturbed by at least one of the RDBV members, implying that the true state will rarely be an outlier in the context of a Talagrand histogram. This combination of under- and over-dispersiveness leads to the bimodal structure observed in Figure~\ref{fig:L96TalHistDA} for RDBVs.

We remark that in the strongly localised case $K=128$, increasing the perturbation size $\delta$ does not mitigate the issue of unreliability. We found that the values of $\delta$ needed to generate a flat Talagrand histogram feature significantly larger forecast errors (not shown). Likewise, improving the accuracy of the observations does not allow for SPBVs and RDBVs to be reliable, but the associated smaller analysis error only causes the poor reliability to occur for an associated smaller optimal perturbation size $\delta$.\\

%We therefore have to conclude that the Talagrand histogram alone is inappropriate to fully assess the reliability of localised BV ensembles.

%In the idealised case, the noise strength $\sigma$ used to generate SPBV ensembles could be tuned to obtain reliable SPBV ensembles. Naively we may expect that increasing $\sigma$ will also lead to more reliable SPBV ensembles in the case when ensembles are generated from the analysis. However, in the strongly localised case $K=128$ increasing $\sigma$ only increases the variance within the localised region and does not affect the components outside the region. As for the RMS forecast error, we find that reliability measures are insensitive to changes of $\sigma$ and reliability cannot be achieved by tuning $\sigma$. We find that for all values of $\sigma$ SPBV ensembles are under-dispersive. 
%%\gaginline{ !!! Maybe $\sigma=1.25$ is too large? Or maybe we need to sample $\sigma$ more finely as indicated by the rapid (in $\sigma$) change of the RMS error when varying $\sigma$. Can we see the Talagrand diagram for some values of $\sigma$ with that transition region in the RMS?} 
%The error-spread ratio and Talagrand histogram are shown in Figures~\ref{fig:L96ErrSprdSigmaDA} and \ref{fig:L96TalHistSigmaDA} respectively for $\sigma = 0.25$, $\sigma = 1.25$ and $\sigma = 5.0$. Again, the error-spread ratio and Talagrand histograms are obtained for the respective optimal values of $\delta$ for which the RMS error attains its minimum value for each ensemble type. The reliability of SPBV ensembles saturates for $\sigma \ge 1.25$ and is more under-dispersive for $\sigma < 1.25$ for both dimension sizes $K$.\\
%

Naively we may expect that increasing $\sigma$ will lead to more reliable SPBV ensembles. As for the RMS forecast error, we find that reliability measures are insensitive to changes of $\sigma>1.25$. The reliability of SPBV ensembles saturates for $\sigma \ge 1.25$ and is more under-dispersive for $\sigma < 1.25$ for both dimension sizes $K$ (not shown).\\
 
%In the idealised case, the noise strength $\sigma$ used to generate SPBV ensembles could be tuned to obtain reliable SPBV ensembles. Naively we may expect that increasing $\sigma$ will also lead to more reliable SPBV ensembles in the case when ensembles are generated from the analysis. However, in the strongly localised case $K=128$ increasing $\sigma$ only increases the variance within the localised region and does not affect the components outside the region. As for the RMS forecast error, we find that reliability measures are insensitive to changes of $\sigma$ and reliability cannot be achieved by tuning $\sigma$ and the reliability of SPBV ensembles saturates for $\sigma \ge 1.25$ and is more under-dispersive for $\sigma < 1.25$ for both dimension sizes $K$ (not shown).\\

We remark that the property of an ensemble to be dynamically adaptive, i.e. their relationship with covariant Lyapunov vectors %(cf. Section~3\ref{sec.CLV}) 
and that they are conditioned on the current state, does not seem to be necessarily promoting improved forecast skill and reliability. In fact, the dynamically non-adapted RDBVs perform better than the dynamically adapted SPBVs with regards to forecast skills for both $K=40$ and $K=128$, and in the weakly localised case $K=40$ they are also slightly more reliable.  

%\begin{figure}
%	\centering
%	\includegraphics[width=19pc]{L96K40ErrSprdSigmaDA.pdf}\\
%	\includegraphics[width=19pc]{L96K128ErrSprdSigmaDA.pdf}
%	\caption{RMS error vs RMS spread for SPBVs with different noise strength $\sigma$ parameterised by increasing lead times from $\tau=0$ to $\tau=5.0$ time units with $\delta = 0.1$. The markers indicate the specific lead times $\tau = 2.0$, $\tau = 3.0$ and $\tau = 4.0$. Top: $K=40$. Bottom: $K=128$.}
%	\label{fig:L96ErrSprdSigmaDA}
%\end{figure}
%
%\begin{figure}
%	\centering
%	\includegraphics[width=19pc]{L96K40TalHistSigmaDA.pdf}\\
%	\includegraphics[width=19pc]{L96K128TalHistSigmaDA.pdf}
%	\caption{Talagrand histograms for SPBVs with differing choices of $\sigma$ for $\delta = 0.1$. Top: $K=40$. Bottom: $K=128$.}
%	\label{fig:L96TalHistSigmaDA}
%\end{figure}

%%%%%%%%%%%%%%%%%%%%%%%%%%%%%%%%%%%%%%%%%%%%%%%%%%%%%%%%%%%%%%%%%%%%%

\section{Discussion and outlook}
\label{sec.summary}

We have explored the framework of stochastically modified bred vectors, developed originally for multi-scale systems in \cite{GigginsGottwald19}, for systems without scale separation. We considered two stochastic modifications, SPBVs which preserve any eventual localisation of the their parent BVs and their spatial correlation structure, and RDBVs which do not do so. SPBVs were constructed to sample the probability density function conditioned on the current state whereas RDBVs are not conditioned on the current state but may evolve into future states which do not reliably estimate the probability density function at a given lead time. The difference in construction renders SPBVs dynamically adapted in the sense that they project onto dynamically relevant covariant Lyapunov vectors whereas RDBVs are not dynamically adapted. Using numerical simulations of the single scale Lorenz 96 model we have shown that SPBVs and RDBVs successfully mitigate the collapse to a single ensemble member of classical BVs with significantly increased ensemble dimension for perturbation sizes $\delta$ in the range of typical analysis errors. Related to this, the forecast skill - as measured by the RMS error -- and the ensemble reliability -- probed by the error-spread ratio and the Talagrand diagram -- are markedly improved by the stochastic modifications. 

We identified the property of localisation of fast growing perturbations which is often observed in spatially extended systems to be a crucial aspect for the performance of stochastically modified BVs. Whereas localisation is advantageous to condition on the current state, it is detrimental in allowing the ensemble to perturb spatial regions of non-vanishing analysis error which are outside the localised region. This causes SPBVs to be under-dispersive (independent of the noise strength). RDBVs exhibit the better forecast skill, despite not being dynamically adapted. In the weakly localised case, RDVs and SPBVs perform equally well, and behave (per construction) statistically similarly, and both ensembles significantly improve the forecast skill and reliability of classical BV ensembles. Our results suggest that the applicability of SPBVs to single-scale systems is limited to situations with small degree of localisation.\\

To counteract the detrimental effect of localisation in SPBVs one could apply additive noise at all sites outside the active localised region, similar to the method proposed in \cite{GreybushEtAl13}. We tried this in the L96 model but did not find that it overcame the problem. The level of noise required to account for the analysis error was found to be such that the noise to BV-signal ratio was too large and the perturbation would be close to a Gaussian random perturbation. This may be though an artefact of the L96 model and the addition of spatially homogenous noise on SPBVs may still mitigate against the problem of localisation in more complex models. \\  

We increased the diversity by introducing stochasticity directly to the bred vectors. Diversity may also be introduced stochastically by adding noise to the evolution equations generating the bred vectors. This can be done in a dynamically consistent way in the context of multi-scale dynamics (see for example \cite{GottwaldEtAl17} and references therein). For multi-scale dynamics exhibiting rapid regime transitions it was moreover shown in \cite{MitchellGottwald12b,GottwaldHarlim13} that stochastically parametrised forecast models for the slow variables significantly improve the analysis of an ETKF as well as the ensemble's reliability. It would be interesting to see if such stochastically perturbed dynamical models can also be used to improve the diversity of bred vector ensembles.\\

%Firstly, perturbing SPBVs by additive noise would case SPBVs to lose their ability to project strongly onto Lyapunov vectors and sample local growing modes. Additive noise causes the SPBVs to lose dynamical consistency as we would be perturbing regions of unlikely fast error growth. Secondly, for additive noise to be effective, it would have to be on the scale of the expected analysis error. Additive noise at a smaller scale will still cause the ensemble to be under-dispersive by a similar argument as previously discussed for SPBVs, whereas much larger additive noise will cause over-dispersiveness. However, additive noise on the size of the typical analysis error is essentially equivalent to sampling $\P_a$, which could be done more effectively by the respective ensemble data assimilation method used to generate the analysis. Moreover, the underlying BV perturbation would be ''swamped" by noise of this size and thus we would lose all information provided by the BV. We note that this analysis is only valid in the context of the L96 system, and the addition of small additive noise could prove beneficial for other dynamical systems not considered here.\\

We would like to stress that our work only considers bred vectors here as a method for probabilistic forecasting and is concerned with improving the breeding method. We do not attempt to compare different methods such as ensemble Kalman filter ensembles, singular vectors and other methods, and to determine their individual merits. We used here as a reference ETKF ensembles with much larger ensemble dimension than the bred vector ensembles. The ETKF ensembles perform very well in the L96 model, providing superior forecast skill and reliability while also being dynamically adaptive. However, which ensemble method performs optimally as a forecast ensemble is in fact situation dependent, as pointed out, for example, recently by \cite{OKaneEtAl19}. The authors found that in a coupled atmosphere ocean model forecast ensembles initialised using bred vectors with perturbation sizes tuned to capture the tropical Pacific thermocline variability, are best suited for ENSO forecasting, compared to ensembles initialised from ETKFs.\\ 

%We caution the reader that one cannot simply extrapolate the bad performance of BVs observed here in the setting of the L96 toy model to an operational setting. In realistic operational forecasting situations, uncertainty in saturated sub-synoptic processes such as convective events often generate sufficient variability in the synoptic scales, and thereby prevent BVs from collapsing onto a single BV \citep{TothKalnay97}. However, our work shows how to improve BVs and mitigate against under-dispersive BV ensembles \citep{Palmer18} without significant additional computational cost, and our work shows that caution needs to be taken in case of strongly localised error growth in situations without strong scale separation. 
%\textcolor{blue}{To expand on the limitations of the L96 model it has to be noted that the L96 model does not feature regime transitions between meta-stable states as observed in the atmosphere. Such transitions often occur on fast time scales with rapid error growth. In such situations the ensemble collapse of classical BVs to a single ensemble member may be exacerbated. It is, however, not clear which os the two stochastic modifications, SPBV or RDBV, are better suited to mitigate against the collapse and which of the two modifications will form the more reliable ensemble.} 

We caution the reader that one cannot simply extrapolate the performance of BVs and their stochastic modifications, SPBV and RDBV, observed in the setting of the L96 toy model to an operational setting. Realistic geophysical fluid models involve the intricate interplay of various processes running on numerous moderately separated time-scales with varying degrees of localisation of error growth, and may exhibit regime transitions between meta-stable states. In realistic operational forecasting scenarios, uncertainty in saturated sub-synoptic processes such as convective events often generate sufficient variability in the synoptic scales, and thereby prevent BVs from collapsing onto a single BV \citep{TothKalnay97}. On the other hand, regime transitions often occur on fast time scales with rapid error growth, potentially exacerbating the ensemble collapse of classical BVs to a single ensemble member. Our work shows how to mitigate against potential under-dispersive BV ensembles \citep{Palmer18} without significant additional computational cost. In particular, our work here on single-scale dynamics and our previous work on multi-scale dynamics \citep{GigginsGottwald19} have identified the degree of time-scale separation and the degree of localisation as key to the performance of the proposed stochastic modifications; whereas localisation of error growth is crucial for the good performance of SPBVs in multi-scale dynamics, caution needs to be taken in the case of strongly localised error growth in situations without strong scale separation. The respective performance of SPBVs and RDBVs in an operational setting or in other realistic geophysical fluid flow applications will depend on the situation-dependent interplay of (moderate) time-scale separation and localisation.

\section*{{\bf{Acknowledgements}}}
BG thanks Diego Paz\'o and Juanma L\'opez for stimulating discussions and for their hospitality. BG acknowledges the support of an Australian Postgraduate Award. GAG acknowledges support from the Australian Research Council, grant DP180101385.

\end{document}